\documentclass[lettersize,journal]{IEEEtran}
\usepackage{amsmath,amsfonts}
\usepackage{amsmath}
\usepackage{algpseudocode}
\usepackage{algorithm}
\usepackage{algpseudocode}
\usepackage{array}
\usepackage[caption=false,font=normalsize]{subfig}
\usepackage{textcomp}
\usepackage{stfloats}
\usepackage{url}
\usepackage{verbatim}
\usepackage{gensymb}
\usepackage{graphicx}
\usepackage{cite}
\usepackage[export]{adjustbox}

\usepackage{pifont}
\newcommand{\cmark}{\ding{51}}%
\newcommand{\xmark}{\ding{55}}%

\usepackage[table,xcdraw]{xcolor}
\usepackage{multirow}
\usepackage{ifthen}
\usepackage{tikz}
\usepackage{tikzsymbols}
\usetikzlibrary{positioning, shapes.geometric, arrows, automata, spy}
\usepackage{wasysym}
\usepackage{scalerel}
\usepackage{tikz}
\usetikzlibrary{svg.path}

\definecolor{color1}{RGB}{223, 225, 255}
\definecolor{color2}{RGB}{0, 0, 160}
\definecolor{color3}{RGB}{51, 51, 51}
\definecolor{color4}{RGB}{225, 255, 233}
\definecolor{color5}{RGB}{0, 64, 0}
\definecolor{color6}{RGB}{255, 216, 176}
\definecolor{color7}{RGB}{0, 0, 0}
\definecolor{color8}{RGB}{254, 226, 235}
\definecolor{color9}{RGB}{255, 138, 138}
\definecolor{color10}{RGB}{179, 254, 174}
\definecolor{color11}{RGB}{251, 252, 207}
\definecolor{color12}{RGB}{233, 206, 22}
\definecolor{color13}{RGB}{255, 255, 255}
\definecolor{color14}{RGB}{255, 179, 178}
\definecolor{color15}{RGB}{162, 177, 195}
\definecolor{color16}{RGB}{138, 138, 138}

\algnewcommand\algorithmicforeach{\textbf{foreach}}
\algdef{S}[FOR]{ForEach}[1]{\algorithmicforeach\ #1\ \algorithmicdo}

\usepackage{hyperref}
\usepackage{orcidlink}

\begin{document}

\title{SALSy: Security-Aware Layout Synthesis}

\author{Mohammad Eslami\orcidlink{0000-0001-7200-3655}, ~\IEEEmembership{Graduate Student Member,~IEEE,}
Tiago Perez\orcidlink{0000-0001-6006-1938}, ~\IEEEmembership{Graduate Student Member,~IEEE,}
Samuel Pagliarini\orcidlink{0000-0002-5294-0606}, ~\IEEEmembership{Member,~IEEE}% <-this % stops a space
\thanks{This work was partially supported by the EU through the European Social Fund in the context of the project “ICT programme".}
\thanks{M. Eslami, T. D. Perez, and S. Pagliarini are with the Department of Computer Systems, Centre for Hardware Security, Tallinn University of Technology (TalTech), 12618, Tallinn, Estonia (e-mail: mohammad.eslami@taltech.ee; tiago.perez@taltech.ee; samuel.pagliarini@taltech.ee)}}% <-this % stops a space
%\thanks{Manuscript received April 19, 2021; revised August 16, 2021.}}% <-this % stops a space
%\thanks{Manuscript received April 19, 2021; revised August 16, 2021.}}



\maketitle

\begin{abstract}
Integrated Circuits (ICs) are the target of diverse attacks during their lifetime. Fabrication-time attacks, such as the insertion of Hardware Trojans (HTs), can give an adversary access to privileged data and/or the means to corrupt the IC's internal computation. Post-fabrication attacks, where the end-user takes a malicious role, also attempt to obtain privileged information through means such as fault injection and probing. Taking these threats into account and at the same time, this paper proposes a methodology for Security-Aware Layout Synthesis (SALSy), such that ICs can be designed with security in mind in the same manner as power-performance-area (PPA) metrics are considered today, a concept known as security closure. Furthermore, the trade-offs between PPA and security are considered and a chip is fabricated in a 65nm CMOS commercial technology for validation purposes -- a feature not seen in previous research on security closure. Measurements on the fabricated ICs indicate that SALSy promotes a modest increase in power in order to achieve significantly improved security metrics.

%Although outsourcing IC fabrication brings numerous advantages to the companies, there are some threats made since some of the post-design steps in the IC supply chain are considered to be untrusted. 

%Hence, protecting the physical layout passed to the foundry to be fabricated is crucial for ensuring the security of the IC design and preventing unauthorized access to sensitive information. Unauthorized access to the physical layout can lead to the theft of intellectual property, the introduction of malicious hardware (known as hardware Trojans), and other security risks. Therefore, IC design houses must take steps to secure their physical layout. 
%Some studies only address the mentioned vulnerabilities without implementing their work on a fabricated chip, and the lack of a practical silicon demonstration is sensed to validate the efficiency of the introduced approaches in real chips.
%This paper presents a thorough study of the measurements and techniques that can be used to harden the layout against potential threats and attacks during the fabrication and post-fabrication phases. Unlike previous works that only aimed for security closure, we validate our techniques by fabricating a chip with security properties in a commercial 65nm CMOS technology.
\end{abstract}

%The trend of outsourcing components and services from around the world to create a complete product, which is known as globalization, is becoming more popular in the IC supply chain. 


\section{Introduction} \label{sec1}
\IEEEPARstart{G}{lobalization} of the Integrated Circuit (IC) supply chain has brought several benefits to IC vendors, including increased efficiency, cost savings, and access to specialized fabrication. The IC supply chain, today, is a complex network of entities involved in the processes of designing, manufacturing, testing, distributing, and marketing ICs. The limited availability of advanced silicon manufacturing sites, i.e., foundries, is the embodiment of the process of globalization. With the cost to establish a foundry in the range of billions of dollars, only a few companies are left in the competition for cutting-edge chip manufacturing. Hence, IC design companies can avoid these capital expenses by outsourcing the fabrication process and instead concentrating on their core skills, such as designing their specific ICs and the systems around them \cite{tsmc}. %here a reference to Apple would fit well.
 This arrangement in which IC design companies are fabless has been the norm for many years, and it is sustained by the significant investments semiconductor foundries make in R\&D.

%Constructing and maintaining a semiconductor fabrication facility is a highly capital-intensive process that requires sizable investments in infrastructure, labor, and machinery.

%Another reason for outsourcing IC fabrication is to gain access to specialized technologies from other companies. In order to develop and improve their fabrication processes, semiconductor foundries make significant investments in research and development. As a result, they are able to produce ICs efficiently in terms of performance, power usage, and reliability. Companies can take advantage of these specialized technologies and expertise without incurring the expenses of developing them internally by outsourcing their fabrication to these foundries.

However, globalization has also created new security challenges. In the fabless model, the foundries are considered untrusted since design houses have no ownership or oversight claims over them. Hence, IC design houses should seek to protect their designs (layouts) against potential adversaries located in the untrusted foundries \cite{HwSOverview, PrimeronHS, ProtectTrj, DETERRENT}. Such adversaries could perform IP theft, IC overproduction, many forms of reverse engineering, and also compromise the IC's functionality or reliability \cite{breakingSilic, chipREsurvey}. Untrusted foundries may introduce malicious hardware, known as Hardware Trojan (HT), into the IC design. HTs can compromise device functionality or security \cite{htsurv, tehrani10}. For instance, a foundry may introduce a backdoor into an IC that allows an attacker to remotely control the device, steal sensitive data, or inject malicious code \cite{backdoor}. 

Furthermore, there are many other threats beyond fabrication-time attacks. The finalized IC, once being available to a malicious end-user from the open market,  may be targeted by an adversary through fault injection \cite{chipFI, chipREsurvey, ChipFI2}. In this type of attack, the adversary tries to compromise the security of the chip by injecting different types of faults into it. 

Another post-fabrication time attack is probing, in which the attacker tries to gain unauthorized access to the internal data of a chip by performing physical probing \cite{probing, tehraniFSP}. This attack is mostly performed to extract sensitive data from within the chip, such as cryptographic keys or other proprietary information. Such attacks are even more relevant in dependable/critical applications~\cite{ChipFI2}.

With these concerns in mind, the notion of security closure \cite{eslami23} has been pursued by hardware security researchers. It involves accepting certain overheads in terms of power-performance-area (PPA) to achieve heightened security measures, which aim to minimize vulnerabilities and potential attack surfaces. The primary objective is to create trustworthy and robust ICs capable of withstanding potential security breaches and ensuring their reliable performance.

This paper presents a methodology %n IC design flow
for security closure consisting of different techniques. Our proposed flow, Security-Aware Layout Synthesis (SALSy), is generic and can be adopted in any layout, regardless of size, type, or technology. The main contributions of this work are as follows:

\begin{itemize}
    \item Providing a generic approach for enhancing the security of designs during physical synthesis against multiple threats: (i) HTs, (ii) fault injection, and (iii) probing.
    \item Prototyping a chip in a commercial 65nm CMOS technology to validate SALSy in silicon.
    \item Comparing the use of commercial libraries and Process Design Kits (PDKs) with open-source ones in order to highlight the limitations and restrictions of using open-source PDKs for security research.
    \item Making the scripts readily accessible to the public to empower the research community to comprehensively verify and validate the techniques presented in this study. Furthermore, the scripts operate within a commercial physical synthesis tool, assuring that SALSy is compatible with current industry practices. 
\end{itemize}

The rest of the paper is organized as follows: Section \ref{sec2} provides background on major threats against the ICs after being sent for fabrication. In Section \ref{sec3}, the motivation of this work is explained, followed by the details and different techniques we used to enhance the security of the designs using open-source PDKs in Section \ref{sec3a}. Section \ref{sec4} points out the differences between commercial PDKs and open-source ones. The experimental results from the chip are presented in Section~\ref{sec5}, followed by a discussion in Section~\ref{sec6}. Finally, Section \ref{sec7} concludes the paper.

%MISSING
%length
\section{Background} \label{sec2}
\begin{figure*}[tb]
    \centering
    \includegraphics[width=0.85\textwidth, trim=.1cm 18pt .1cm 18pt, clip]{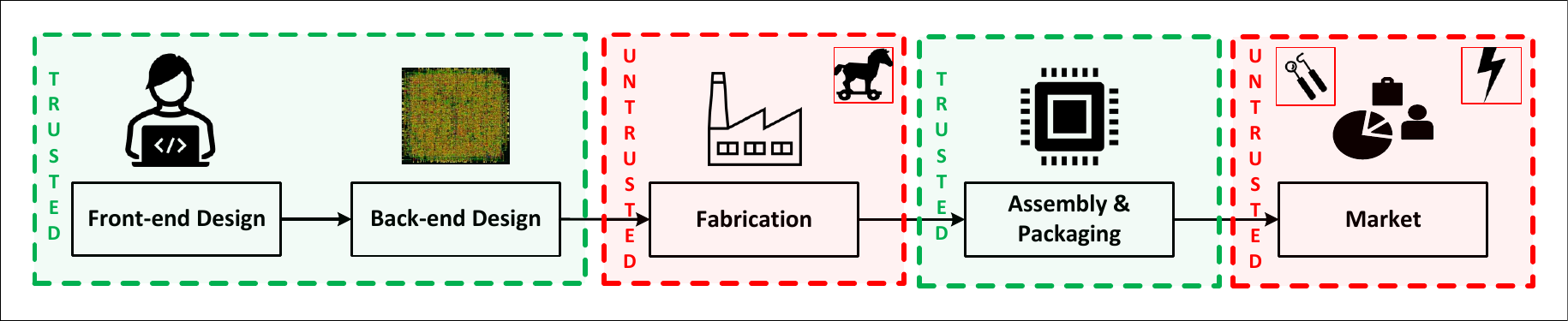}
    \vspace{-3mm}
    \caption{Different threats in the IC supply chain: HTs are a fabrication-time threat, FSP/FI are post-fabrication threats.}
    \label{fig:threats}
\end{figure*}

The IC supply chain encompasses various stages, from design and manufacturing to distribution and deployment, as depicted in Fig. \ref{fig:threats}. Each stage presents unique security challenges and potential threats. %Here are the different threats typically encountered in different stages of the IC supply chain:

Once the chip is sent for fabrication, the design team has no control over it. Hence, considerations must be taken into account to protect the chip against mentioned threats in the fabrication and post-fabrication stages. In this work, we focus on three major post-design threats: \textit {\textbf{Hardware Trojans, Fault Injection, and Probing.}} %More details about the threat model are brought in Section \ref{sec3}. 
In the following, we provide more details about each of these threats.

\subsection{Hardware Trojans}

As mentioned, a HT is a malicious modification to an IC that can cause harm to the system it is embedded in and can remain undetected for long periods of time. Its purpose is to modify the behavior of the IC in a way that benefits the attacker, harms the user, leaks sensitive information, or causes the IC to fail under specific conditions \cite{htsurv, TrjClass}. HTs can be introduced into an IC in several ways, such as by modifying the layout of the design or by manipulating the fabrication process to introduce defects \cite{trojanLessons}. The HT's activation mechanism is called `trigger' -- the event or condition that initiates the attack. The trigger could be a specific input, a particular sequence of operations, or even a specific date or time \cite{tehrani10}.

The specialized literature classifies HTs into two main categories: functional and parametric. Functional HTs alter the functionality of a circuit, while parametric HTs change the circuit's performance parameters. This work focuses primarily on additive functional HTs. % rather than subtractive or substitution ones. 
Additive HTs involve the insertion of additional malicious components into an IC. Conceptually, the HT logic can replace filler cells in a finalized layout \cite{Alex, tiago_tcad}.

\subsection{Fault Injection}

Fault injection has emerged as a potent attack vector that adversaries can exploit to compromise the security and reliability of ICs \cite{chipFI}. Fault injection attacks involve intentionally disturbing digital circuits to disrupt their regular operation, manipulate data, or bypass security mechanisms \cite{laserFI, laserFI2}. Adversaries can exploit various fault injection techniques and target specific vulnerabilities to achieve their malicious objectives \cite{chipFI, ChipFI2}. By tampering with supply voltages, clocks, or even electromagnetic fields, attackers can induce faults that lead to system failures, unauthorized access, or information leakage. Circuits that implement cryptographic operations are particularly sensitive to such attacks. Mechanisms exploited in fault injection attacks include (but are not limited to):

\textit {Timing Manipulation:} Fault injection attacks often target the precise timing of digital circuits, exploiting vulnerabilities in clock signals, synchronization mechanisms, or critical timing paths to cause errors or disrupt the intended operation.

\textit {Power and Voltage Manipulation:} Tinkering with power supply levels, injecting voltage spikes, or inducing power glitches can lead to circuit malfunction.

%\textit {Security Vulnerabilities:} Fault injection attacks can bypass security mechanisms such as encryption or authentication protocols by manipulating cryptographic operations, compromising the integrity and confidentiality of sensitive data.

\subsection{Probing}
This type of attack involves physical access and measurement of internal signals within a digital circuit \cite{probing, backside}. Adversaries employ various methods, such as utilizing oscilloscopes, logic analyzers, or even direct physical contact, to capture and analyze signals. This technique allows attackers to extract sensitive information, such as encryption keys, proprietary algorithms, or critical data, by bypassing traditional security mechanisms. Probing is often performed in two ways:

\textit {Front-side Probing:} This method allows unauthorized access to the internal data of an IC by making electrical connections with specific points on the surface of the chip (metal interconnects and active components) using specialized equipment like microprobes or needles. By reading the internal signals, an attacker can potentially extract sensitive information, including cryptographic keys or proprietary data.
%In digital circuits, unintended signal leakage can occur due to various factors, including electromagnetic radiation, power consumption, or electromagnetic coupling. Attackers can exploit these leakages to deduce valuable information about the internal operations and sensitive data being processed within the circuit.

\textit {Back-side Probing:} This method involves physically accessing the back or underside of the chip to obtain internal data, requiring the removal of the protective packaging to expose the silicon die. This more invasive and destructive technique enables direct access to the chip's internal circuitry, providing valuable insights into its behavior.
%Physical tampering, such as inserting probes directly into the circuit or modifying the circuit layout, 
%can provide attackers with unrestricted access to internal signals, enabling them to extract sensitive information or manipulate the circuit's behavior.
%added here

Front-side probing and fault injection threats share similarities, as both require physical access to the IC. Combating one often involves addressing the other, as their countermeasures overlap, including secure layout design. From this point onward, we will refer to these threats as one under the abbreviation \textbf{FSP/FI}.

\subsection{Related Works}

While numerous works focus on mitigating post-design time attacks through methods like shielding to protect specific chip areas, only a limited number of studies propose defensive techniques targeting multiple threats during the physical synthesis phase -- the final step before chip fabrication. Notably, the vast majority of proposed techniques have not been validated in silicon. This is particularly true for HT prevention techniques.

Defensive techniques proposed against HT insertion are mainly based on the idea of increasing the design's density, thus  reducing the available space within the design where an attacker could potentially embed malicious logic. 

In \cite{second}, a locking technique is introduced during physical synthesis to enhance security and bridge layout gaps. Despite its resilience to various attacks, its practicality is debatable as meeting timing constraints remains uncertain for many benchmarks.  In \cite{bisa}, authors propose populating unused spaces with functional cells, creating an independent combinational circuit for post-fabrication testing against cell modifications. A limitation of this approach is achieving a high occupation ratio while maintaining the design routable. To address this limitation, \cite{papa} prioritizes filling gaps that could be used for HT insertion. However, this approach may alter initial routing, potentially defrauding critical paths due to rerouting.

In \cite{JohannICCAD}, another selective approach places sensitive logic in denser layout areas and steers gaps around less sensitive regions. However, the user's control over placement can be limited, particularly when using commercial CAD tools. In \cite{ASSURER}, authors address this by suggesting a placing refinement technique to segment large unused layout spaces. Yet, this approach faces a significant vulnerability where attackers can reverse the refinement, creating optimal zones for their malicious logic. 

Countermeasures against FSP/FI are principally based on the idea of covering security-sensitive elements with different metal layers since the attacker usually uses laser or ion beams to attack the chip. In \cite{JohannICCAD}, a method is introduced to route sensitive wires beneath regular ones and widen non-sensitive wires. While this incurs substantial overhead in routing resources, it does not assure wire protection, especially against advanced attack techniques. To tackle this, \cite{ASSURER} proposes maximizing sensitive element coverage by rerouting nets and utilizing available design tracks. However, this approach's applicability is uncertain, particularly as the work uses an open-source PDK. %due to potential Design Rule Check (DRC) violations during routing, particularly as the work uses an open-source PDK.

Shielding offers another avenue to safeguard sensitive elements. In \cite{ter}, guard wires are introduced to shield sensitive nets, yet the focus remains on fabrication-time attacks. Alternatively, \cite{tehraniFSP} presents an anti-probing physical approach via added steps in the synthesis flow. However, this method's efficiency faces challenges with increased sensitive element numbers, and it entails area and power overheads.

\subsection{Threat Model}

A vital step to implementing countermeasures against different attacks is identifying and focusing on the exact attacker's capabilities, intentions, and limitations -- i.e., establishing a coherent threat model. As shown in Fig. \ref{fig:threats}, we assume all design stages (i.e. front-end, back-end) in the design house are performed in a %safe
trusted environment. We also assume the packaging-related activities are performed in a trusted setting.

Still referring to Fig.~\ref{fig:threats}, the first threat we consider in this work is that of a fabrication-time HT insertion. We assume that a rogue engineer inside the foundry has the ability to insert HTs into the finalized layout, and he/she is capable of using any type of advanced CAD tool. Since the adversary works for the foundry, he has access to the PDK. In our threat model, we specifically focus on additive HTs. %, and not the substitution/subtractive ones in our threat model. 

When the already packaged chip reaches the market as a ready-to-use product, we consider the end-user to be a potential adversary. In line with the specialized literature, we assume he/she is capable of using advanced machines such as laser or focused ion beam generators, as well as precise measurement devices to perform fault injection and/or probing. In both cases, the goal of the attacker is to extract sensitive information out of the chip, either by injecting different types of fault or by creating a cavity through milling to expose the sensitive nets, followed by depositing a conductor in the cavity to form a contact pad on the chip's surface. In this case, an attacker is able to probe the created pad to extract the sensitive data. It should be noted that for the probing, we only consider front-side probing. 

\section{Motivation} \label{sec3}

The initial motivation for this work is coming from the authors' participation in a hardware security contest that aimed to protect designs against post-design threats \cite{contest}. While being challenging, due to some limitations, only open-source libraries and PDKs can be used in these types of contests in order to make sure the contest does not restrict participation due to (lack of) access to realistic industry PDKs. Nevertheless, one should validate the devised techniques against commercial cell libraries, PDKs, and tools, in order to ensure their practicality in real applications.

%(see Section \ref{sec4} for more details). 

\subsection{ISPD'22 Contest: Security Closure of Physical Layouts}

This contest was part of the 2022 edition of the International Symposium on Physical Design (ISPD). The participants of this contest were to focus on enhancing the security of the physical layout of ICs. As security engineers, teams were required to systematically and proactively assess and address vulnerabilities in IC layouts during the design phase, specifically targeting the selected threats of HT insertion, probing, and fault injection.% These threats were chosen as they represent relatively straightforward scenarios directly related to physical design and their defenses. 

\subsubsection{Contest Structure}

To make the participants comfortable to use their own preferred (commercial, open-source, or in-house-developed) tools, a Design Exchange Format (DEF) file, and other related post-synthesis files were given to the teams as the baseline. Teams were then asked to do their best to enhance the security of the provided designs. Proposed solutions could be submitted to an online evaluation system for scoring. The benchmarks chosen for this contest mostly consisted of crypto cores that were divided into alpha-round benchmarks (CAST, Camellia, MISTY, PRESENT, OpenMSP430\_1, and three versions of AES \cite{benchmarks, benchmarks1, benchmarks2}) and final-round benchmarks (SEED, TDEA, OpenMSP430\_2, and SPARX \cite{benchmarks, benchmarks1}). Teams advanced to the final round by submitting valid designs for all alpha-round benchmarks.

\subsubsection{Scoring}

The overall score is calculated using Eq.~\ref{EQ1}. To mimic the real challenges that a security engineer would face during physical synthesis, design quality (i.e., PPA) was considered in the final scores as well. Hence, the participants not only had to consider enhancing the security but also try to maintain the design quality at an acceptable level. The metrics for design quality consisted of equally-weighted distribution of power, performance (in terms of clock frequency), area, and routing quality. The baseline provided by the organizers had a score of 1. Submissions that improve design quality or security would be scored between $[0,1)$, whereas poor submissions would be scored between $(1,\infty$].

\begin{equation}{\label{EQ1}}
    Score = Design Quality \times Security 
\end{equation}

For the security component of Eq.~\ref{EQ1}, equally-weighted metrics for HT insertion and FSP/FI were considered. For FSP/FI,  only single-directional attacks from the front side are considered. Since the Security component is composed of two other metrics, Eq. \ref{EQ1} can be represented as follows:

\begin{equation} \label{eq2}
    Score = des \times ti \times fsp\_fi
\end{equation}

where \emph{des} denotes the design quality, \emph{ti} denotes the HT insertion and \emph{fsp\_fi} is front-side probing and fault injection.

\begin{figure}[tb]
    \centering
    \includegraphics[width=0.4\textwidth]{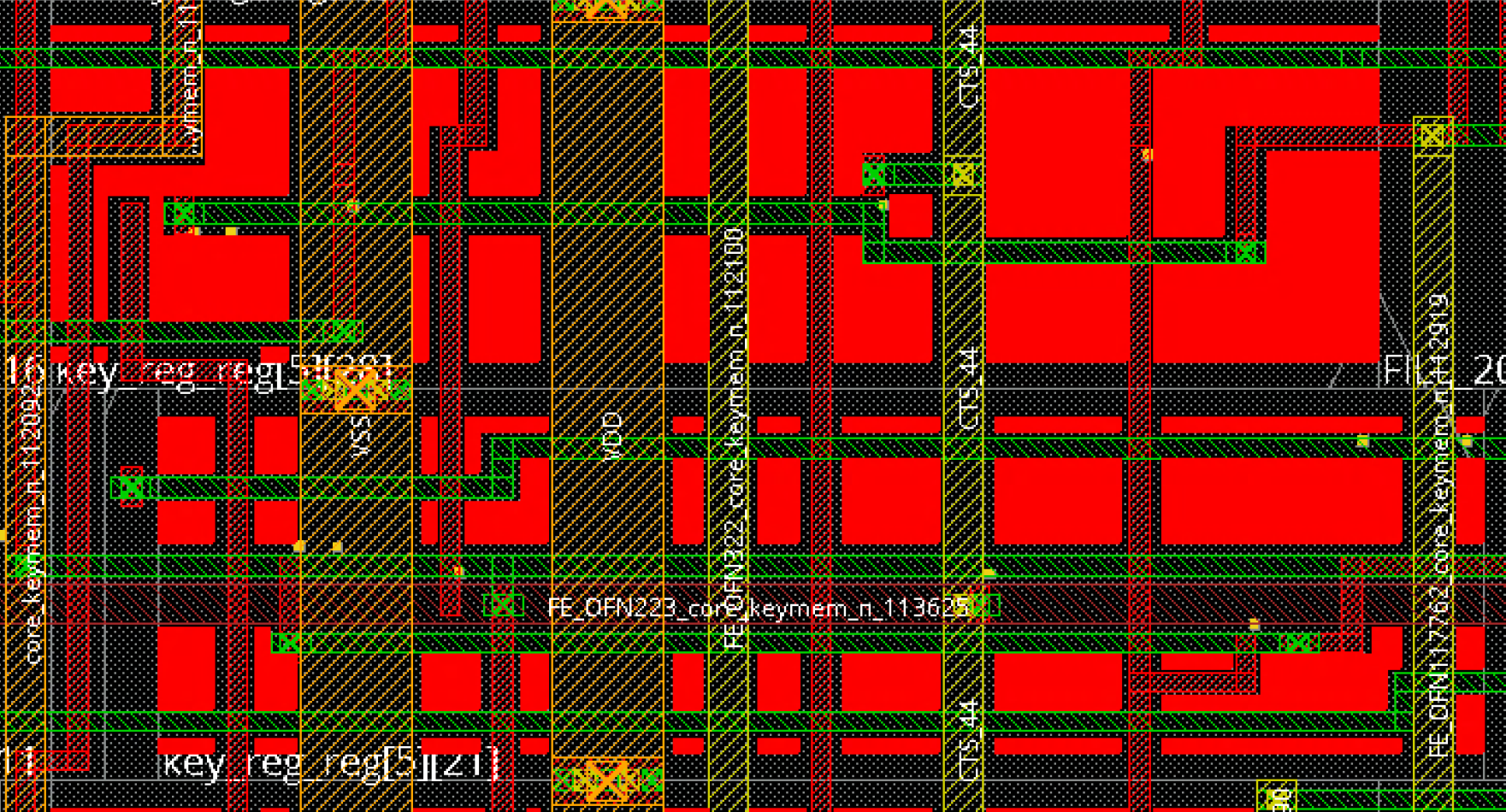}
    % \vspace{-2mm}
    \caption{Example of exposed area (highlighted in red) for cell assets (adopted from \cite{contest}).}
    \label{fig:exposed_area}
\end{figure}

To obtain the security scores for FSP/FI, first, a set of sensitive cells (logical elements) and their related interconnections (wires) were introduced to the participants for each design. The cells are termed as \textbf{\textit{cell assets}} and the wires are called \textbf{\textit{net assets}}. After that, a so-called \textbf{\textit{exposed area}} metric is calculated for each set of cell and net assets in each design. The exposed area is referred to any spatial area of those cell/net assets, whether continuous or fragmented, that is reachable through the metal stack from the top. An example of an exposed area is shown in Fig.~\ref{fig:exposed_area}.

For HTs, an \textbf{\textit{exploitable region}} metric was defined as the set of continuous placement sites\footnote{A placement site represents a predefined valid location where a cell be legally placed; placement sites are typically a function of the standard cell height and the contacted poly pitch.} that are either i) free ii) occupied by filler cells or non-functional cells, or iii) unconnected cells. The criterion for identifying continuous placement sites as an \emph{exploitable region} is met when the number of the sites reaches a minimum threshold of 20. Moreover, free routing tracks around the exploitable region(s) were also considered. The main idea is that an adversary needs placement resources and routing resources to insert an HT successfully. Hence, there should be enough gaps in the layout or some logic that can be easily removed. 

Each of the mentioned parameters in \ref{eq2} has specific weights and is calculated as follows:

\begin{itemize}
    \item For the Trojan insertion (\emph{ti}), 60\% of the score comes from the placement sites of exploitable regions (\emph{ti\_sts}), and the other 40\% comes from the routing resources of exploitable regions (\emph{ti\_fts}).
\end{itemize}

\begin{itemize}
    \item For the front-side probing and fault injection (\emph{fsp\_fi}), the score is equally distributed between the exposed area of the cell assets (\emph{fsp\_fi\_ea\_c}) and the exposed area of the net assets (\emph{fsp\_fi\_ea\_n}).
\end{itemize}

\begin{itemize}
    \item For the design quality (\emph{des}), 10\% of the score is dedicated to the power (\emph{des\_p\_total}), 30\% to the performance (\emph{des\_perf}) in terms of timing violations (if any), 30\% to the area (\emph{des\_area}), and 30\% to  Design Rule Checks (DRCs) (\emph{des\_issues}).
\end{itemize}

\subsubsection{Contest Logistics}

The contest did not restrict the use of any tooling from specific vendors. The PDK/standard cell library of choice is the Nangate 45nm Open Cell Library \cite{nangate} since it is freely available. The metal stacks considered were 6M and 10M, depending on (complexity of) the benchmark.

%material, it has major drawbacks when compared to commercial libraries. More details are brought in Sec. \ref{sec4}.

\section{Security-Aware Layout Synthesis}\label{sec3a}

Let us now introduce SALSy, the main contribution of this work. We will first describe the strategies employed during the contest, and later we will narrow down the strategies only to those practical ones exercised in our tapeout. The strategies and their relative order are shown in Fig.~\ref{fig:salsy}.

We remind the reader that the competition emphasized a balance between security and PPA, as evidenced by the scoring formulas. On the design side, even if we employed customized implementation scripts for each benchmark, our implementations will not be discussed in detail since the scripts implement a fairly traditional flow. In the text that follows, we will focus on the security aspects. 

Since the scoring system considers different metrics for front-side probing/fault injection (\textit{fsp\_fi}) and HT insertion (\textit{ti}), we explain the related techniques separately.

\tikzstyle{shape1} = [trapezium, trapezium left angle=70, trapezium right angle=110, minimum width=6cm, minimum height=1cm, text centered, font=\large, color=color3, draw=color2, line width=1, fill=color1]
\tikzstyle{shape2} = [rectangle, minimum width=5cm, minimum height=1cm, text centered, font=\large, color=color3, draw=color5, line width=1, fill=color4]
\tikzstyle{shape3} = [rectangle, minimum width=5cm, minimum height=1cm, text centered, font=\large, color=color3, draw=color5, line width=1, fill=color4]
\tikzstyle{shape4} = [rectangle, minimum width=5cm, minimum height=1cm, text centered, font=\large, color=color3, draw=color9, line width=1, fill=color8]
\tikzstyle{shape5} = [diamond, minimum width=1cm, minimum height=1cm, font=\normalsize, color=color3, draw=color7, line width=1, fill=color10]
\tikzstyle{shape6} = [diamond, minimum width=1cm, minimum height=1cm, font=\normalsize, color=color3, draw=color12, line width=1, fill=color11]
\tikzstyle{shape7} = [rectangle, minimum width=1cm, minimum height=1cm, text centered, font=\large, color=color3, draw=color7, line width=1, fill=color13]
\tikzstyle{shape8} = [rectangle, rounded corners, minimum width=5cm, minimum height=1cm, text centered, font=\large, color=color3, draw=color2, line width=1, fill=color1]
\tikzstyle{shape9} = [thick, draw=color16, line width=2, ->, >=stealth]

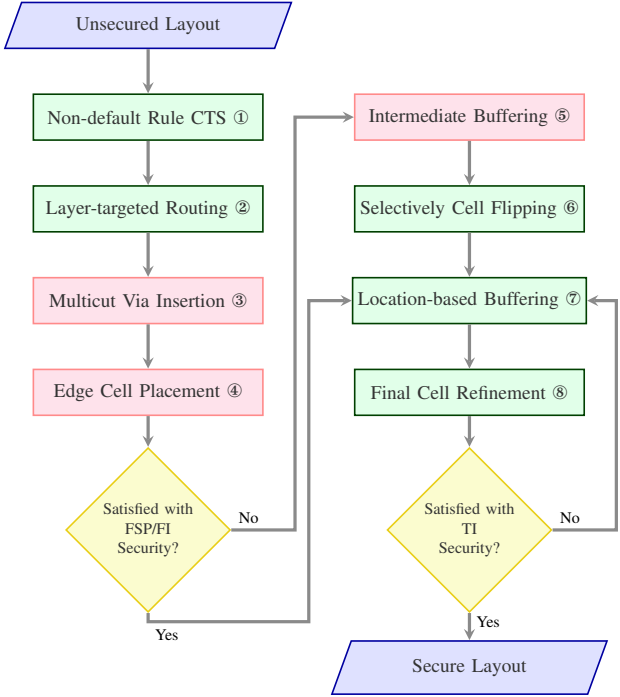
\begin{figure}[tb]
\centering
\begin{adjustbox}{width=0.45\textwidth}
\begin{tikzpicture}[node distance=2cm, on grid, auto]
\node (obj1) [shape1]                                               {Unsecured Layout};
\node (obj2) [shape2, below of=obj1]                                {Non-default Rule CTS \ding{172}};
\node (obj3) [shape4, right of=obj2, xshift=5cm]                    {Intermediate Buffering \ding{176}};
\node (obj4) [shape2, below of=obj2]                                {Layer-targeted Routing \ding{173}};
\node (obj5) [shape2, right of=obj4, xshift=5cm]                    {Selectively Cell Flipping \ding{177}};
\node (obj6) [shape4, below of=obj4]                                {Multicut Via Insertion \ding{174}};
\node (obj7) [shape3, right of=obj6, xshift=5cm]                    {Location-based Buffering \ding{178}};
\node (obj8) [shape4, below of=obj6]                                {Edge Cell Placement \ding{175}};
\node (obj9) [shape2, right of=obj8, xshift=5cm]                    {Final Cell Refinement \ding{179}};
\node (obj10) [shape6, below of=obj9, yshift=-1cm, align=center]    {Satisfied with \\ TI \\ Security?};
\node (obj11) [shape6, below of=obj8, yshift=-1cm, align=center]    {Satisfied with \\ FSP/FI \\ Security?};
\node (obj16) [shape1, below of=obj10, yshift=-1cm]                 {Secure Layout};
\draw [shape9] (obj11) -- node[pos=2]{No} ++(2,0) -| ++(1.2,9)  -- (obj3)  ;
\draw [shape9] (obj11) -- node[pos=2.5]{Yes} ++(0,-2) -| ++(3.5,7)  -- (obj7)  ;
\draw [shape9] (obj1) --  (obj2);
\draw [shape9] (obj2) --  (obj4);
\draw [shape9] (obj4) --  (obj6);
\draw [shape9] (obj6) --  (obj8);
\draw [shape9] (obj8) --  (obj11);
\draw [shape9] (obj3) --  (obj5);
\draw [shape9] (obj5) --  (obj7);
\draw [shape9] (obj7) --  (obj9);
\draw [shape9] (obj9) --  (obj10);
\draw [shape9] (obj10) -- node[pos=2]{No} ++(2,0) -| ++(1.2,5)  -- (obj7)  ;
\draw [shape9] (obj10) -- node[pos=1]{Yes} ++(0,-2) -- (obj16);
\end{tikzpicture}
\end{adjustbox}
 %\vspace{-2mm}
\caption{SALSy framework. Red boxes highlight techniques that are not feasible for the tapeout. Green boxes highlight techniques that can be used in the contest and in the tapeout.}
\label{fig:salsy}
\end{figure}

\subsection{Countermeasures against FSP/FI}

\subsubsection{Non-default Rule Clock Tree Synthesis} The insight of this strategy is to change the default rules for Clock Tree Synthesis (CTS) \footnote{CTS is an essential step in the design process of digital ICs, involving the construction of a network of clock branches to distribute clock efficiently signal across the entire circuit. By carefully constructing the network, observing delay balancing, and skew management, timing can be improved and power consumption can be reduced. CTS routing usually takes precedence and priority over signal routing, which we leverage for security purposes.} in order to cover more assets by enlarging the clock distribution wires. Note that CTS routing utilizes fewer resources than signal routing. Hence, CTS wires can be widened many times more than signal wires. As shown in Fig.~\ref{fig:4techs}\subref{4a}, the enlarged clock tree can significantly cover more exposed areas under it. Quite often the quality of the CTS is improved by using non-default rules.

\subsubsection{Layer-targeted Routing}

Recall that the exposed area metric, which applies to cells and nets, corresponds to the asset area directly accessible from the front side. In the first step, we try to hide the net assets under the other non-asset nets to protect them against FSP/FI, as shown in Algorithm \ref{alg1}. For this purpose, we dedicate the lowest possible metal layers\footnote{Several metal layers form a metal stack, where the lowermost layers are usually the thinnest of them.} to the net assets only (line 3). It should be noted that we use the minimum width for routing these asset-related wires to be later able to `hide' them under the other nets (line 5). 

Next, all remaining non-assets nets are defined to be routed with higher metal layers (line 4). Furthermore, we choose wider width instead of the default one to increase the chance of covering net and cell assets (line 6). If the routing tool fails to route the nets with the modified width or in the preferred metal layer, it will try to route them with the default width and in default metal layers (lines 12-14, 20-22). For the physical synthesis tool utilized, routing constraints are soft constraints. I.e., the tool will try the hardest to follow the constraints; if it fails, the constraints are disregarded. 

\begin{figure*}[ht]
    \centering
    \subfloat[]{\includegraphics[width=0.50\columnwidth, height=0.50\columnwidth]{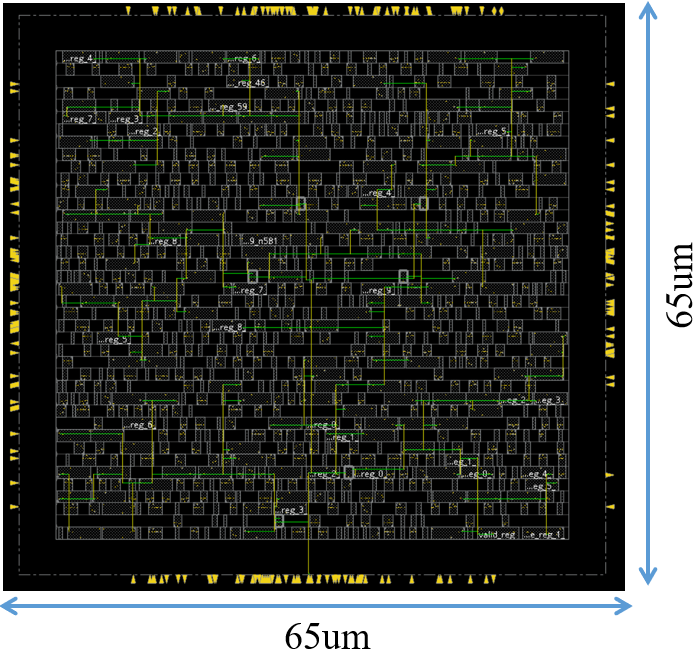}
    \includegraphics[width=0.37\columnwidth]{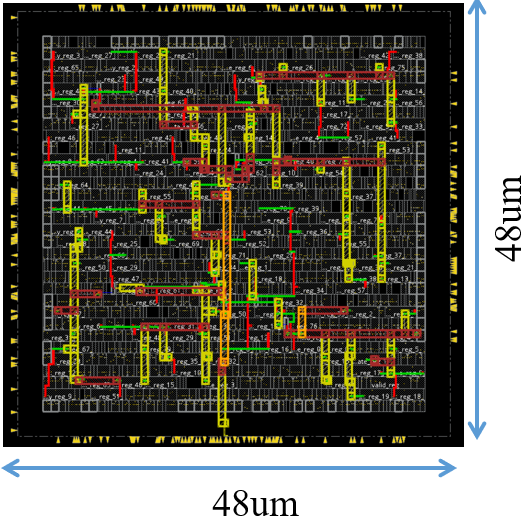} \label{4a}} \hspace{0.05\textwidth}
    \subfloat[]{\includegraphics[width=0.50\columnwidth, height=0.50\columnwidth]{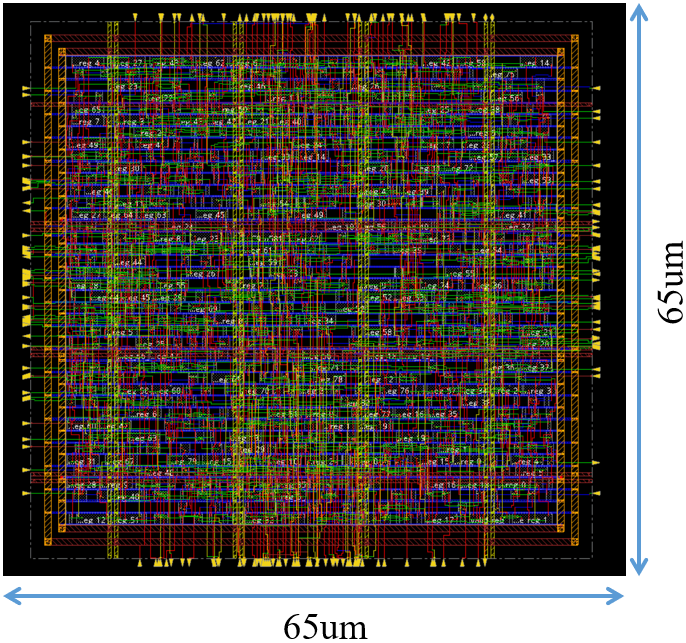}
    \includegraphics[width=0.37\columnwidth]{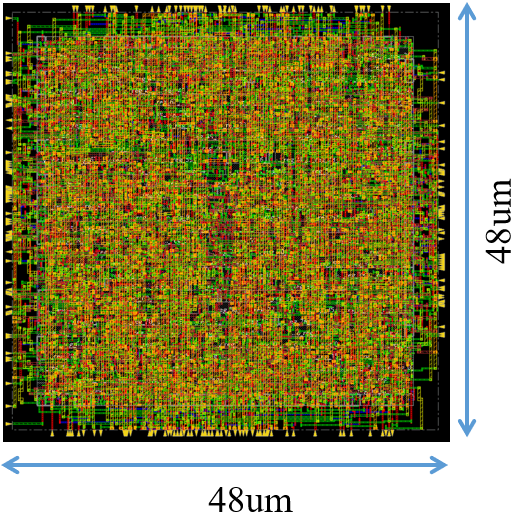} \label{4b}} \hspace{0.05\textwidth}
    \subfloat[]{\includegraphics[width=0.50\columnwidth, height=0.50\columnwidth]{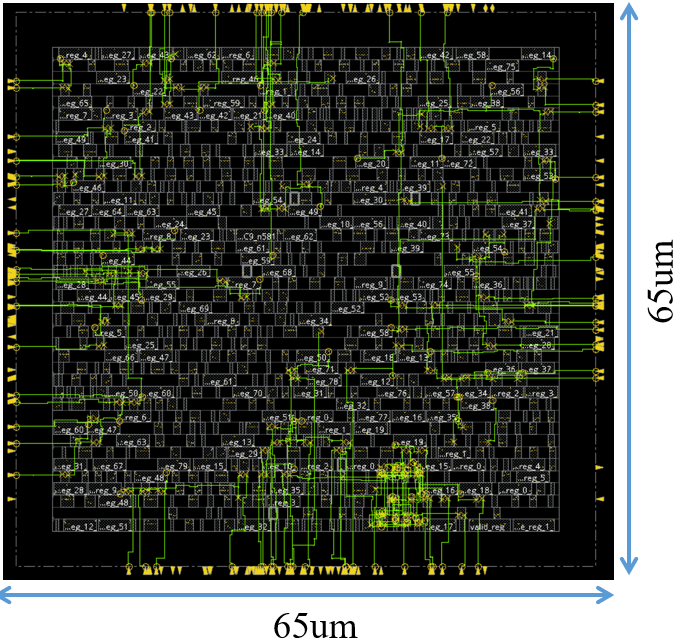}
    \includegraphics[width=0.37\columnwidth]{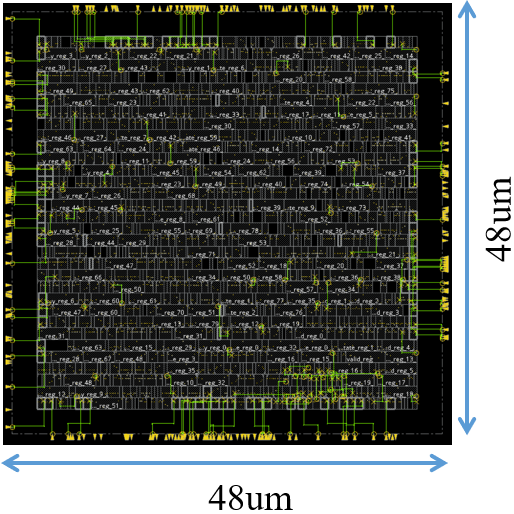} \label{4c}} \hspace{0.05\textwidth}
    \subfloat[]{\includegraphics[width=0.50\columnwidth, height=0.50\columnwidth]{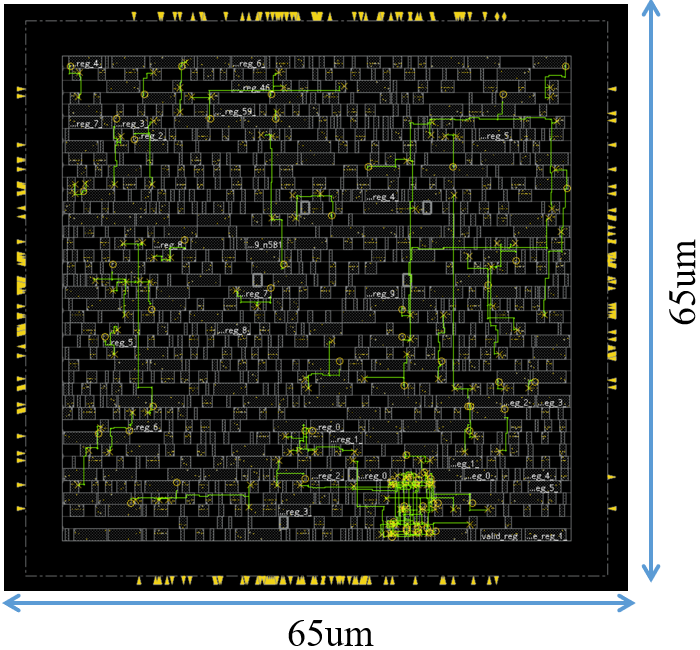}
    \includegraphics[width=0.37\columnwidth]{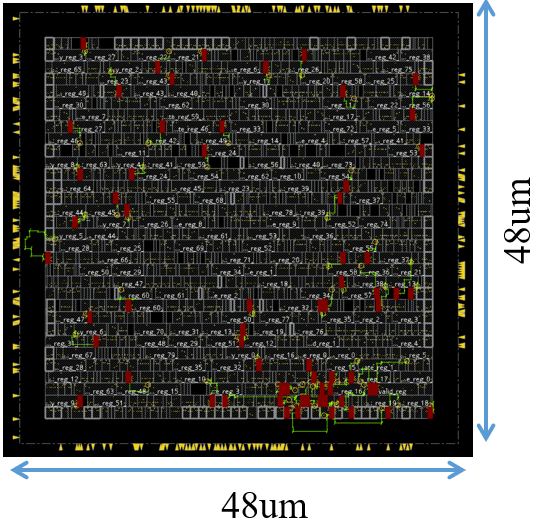} \label{4d}}
    \vspace{-1mm}
    \caption{Different techniques used in SALSy. The design on the left is always the BL variant, and the design on the right is always the SEC variant. a) Non-default Rule CTS, b) Increased congestion by applying Layer-targeted Routing, c) Edge Cell Placement for shortening the long net assets (highlighted in green), and d) Reducing the length of the net assets (highlighted in green) by applying Intermediate Buffering technique (added buffers appear in red).}
    \label{fig:4techs}
\end{figure*}

As an example, in Algorithm \ref{alg1}, we consider M2 and M3 layers for routing only the net assets (line 3), and the higher metal layers M4-M6 are used for non-asset net routing (line 4). In this example, the width of the non-asset nets is 2x wider than the asset nets (line 6), but this number can be increased if more resources are available\footnote{In a real IC, the number of metal layers depends on the technology/metal stack agreed with the foundry. Current technologies often offer 10+ layers.}. After applying this technique, the congestion increases significantly, and therefore, more cell assets and net assets can be protected against FSP/FI, as shown in Fig.~\ref{fig:4techs}\subref{4b}.

\begin{algorithm}[t] 
\caption{Layer-targeted Routing Algorithm} \label{alg1}
\begin{algorithmic}[1]

\State $net\_assets \gets List\_of\_net\_assets$
\State $other\_nets \gets List\_of\_other\_nets$
\State $prf\_lays\_assets \gets [M2,M3] $
\State $prf\_lays\_others \gets [M4,M5,M6] $
\State $width\_for\_assets \gets width(M2) $ \Comment{This value is the minimum width according to the library}
\State $width\_for\_others \gets width(M2)\times2 $ %\Comment{This value can be set in terms of coefficients of minimum width according to the library (here it is defined as 2X)}

\ForEach {$net$ \textbf{in} $net\_assets$}
\State\textbf{assign} \textit{prf\_lays\_assets} \textbf{to} \textit{route\_layer}
\State\textbf{assign} \textit{width\_for\_assets}  \textbf{to} \textit{width\_rule}
\EndFor
\State\textbf{route} \textit{net\_assets} \textbf{with} \textit{width\_ruler} \textbf{in} \textit{route\_layer}

\If{$(route\_err)$} 
\State\textbf{route} \textit{net\_assets} \textbf{with} \textit{default\_rules} 
\EndIf

\ForEach {$net$ \textbf{in} $other\_nets$}
\State\textbf{assign} \textit{prf\_lays\_others}  \textbf{to} \textit{route\_layer}
\State\textbf{assign} \textit{width\_for\_others}  \textbf{to} \textit{width\_rule}

\EndFor

\State\textbf{route} \textit{other\_nets} \textbf{with} \textit{width\_ruler} \textbf{in}  \textit{route\_layer}

\If{$(route\_err)$} 
\State\textbf{route}  \textit{other\_nets}  \textbf{with} \textit{default\_rules} 
\EndIf

\end{algorithmic}
\end{algorithm}

\subsubsection{Multi-cut Via Insertion}

\begin{figure}[ht]
    \centering
    \includegraphics[width=.47\columnwidth]{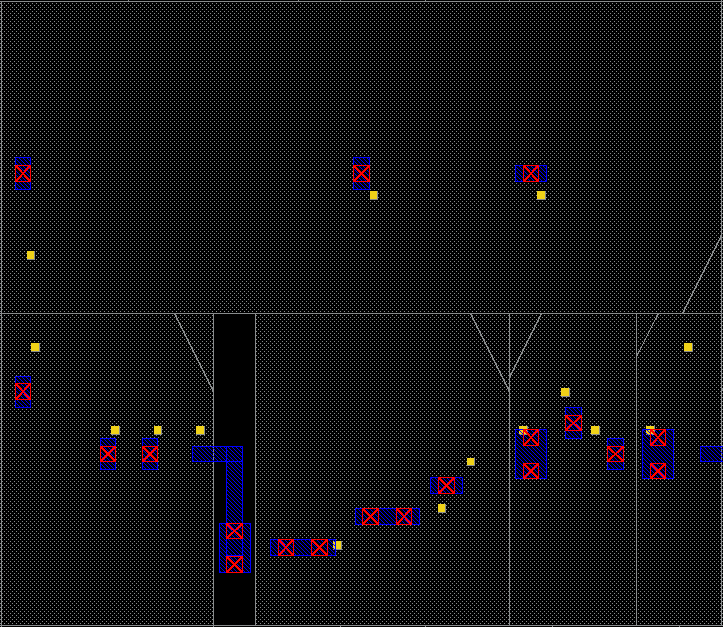}\hspace{0.001\columnwidth}
    \includegraphics[width=.47\columnwidth]{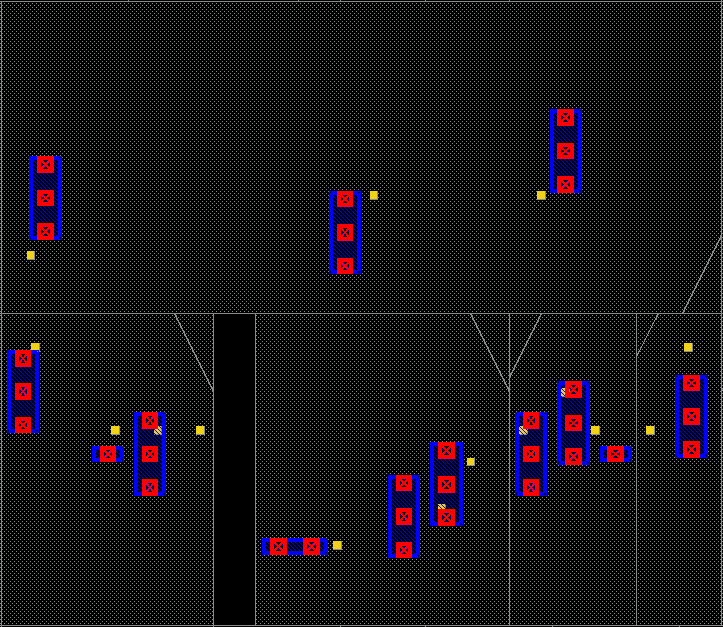}
    \vspace{-2mm}
    \caption{Using the default rules for via insertion (left) and multi-cut via insertion (right) to increase the coverage of cell assets}
    \label{fig:vias}
\end{figure}

A vertical connection named \emph{via} (cut) connects different metal layers in the metal stack of an IC. By default, the physical synthesis tool uses the minimum number of vias and smallest vias available for the connection to optimize routing resource usage and prevent routing congestion. However, our strategy aims to increase the congestion on top of the cell assets to cover them as much as possible. Hence, we use multi-cut vias between M1 and M2 layers. By doing this, a larger piece of metal is routed on the top of the cell assets, improving the coverage, as shown in Fig.~\ref{fig:vias}. The reason that we use multi-cut vias only between M1 and M2 layers is that we do not want to touch the higher metal layer resources, leaving them for signal routing. 

\subsubsection{Edge Cell Placement} 

In some of the benchmarks, it is observed that net assets include long wires that travel a long distance from IO pins to their sinks (see the wires highlighted in green in Fig.~\ref{fig:4techs}\subref{4c}). Hence, we use a technique in which the sink cell connected to the IO pins via net assets is moved to the closest possible position to their driver. In this case, the length of the net assets becomes significantly shorter, which makes it easier to be covered by other nets on upper layers since shorter nets tend to have fewer turns and jogs.

\subsubsection{Intermediate Buffering}

The previous technique for shortening net assets only works for those long wires connected to the IO pins. However, net assets are not always connected to IO pins, and it is challenging to protect them due to their long length. In this case, when both driver and sink are inside the core area (the region where all the cells are placed in), we put buffer(s) in between them to reduce the length of the long net assets, as shown in Fig. \ref{fig:4techs}\subref{4d}. It should be noted that buffer insertion might significantly impact the design's timing and power consumption. Hence, we consider this technique iteratively with multiple checkpoints. If the buffer insertion introduces a timing-violating path, that buffer can be removed, and the circuit goes back to its previous state without the violation.

\subsubsection{Selectively Cell Flipping}

\begin{figure}[tb]
    \centering
    \begin{tikzpicture}[spy using outlines={circle,black!30,magnification=4,size=1.4cm, connect spies}]
    \node {\includegraphics[width=.45\columnwidth, height=.45\columnwidth]{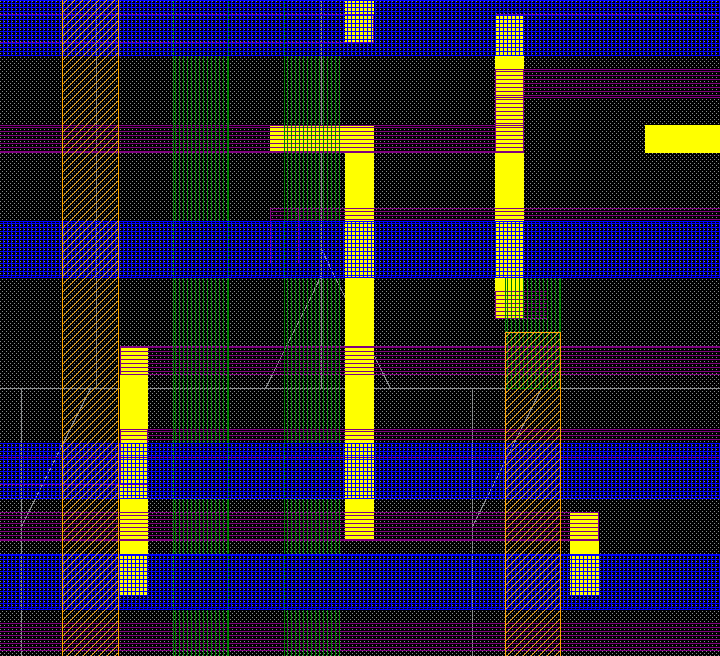}};
    \spy on (0,0.12) in node [left] at (-.55,1.2);
    \end{tikzpicture}
    \begin{tikzpicture}[spy using outlines={circle,black!30,magnification=4,size=1.4cm, connect spies}]
    \node {\includegraphics[width=.45\columnwidth, height=.45\columnwidth]{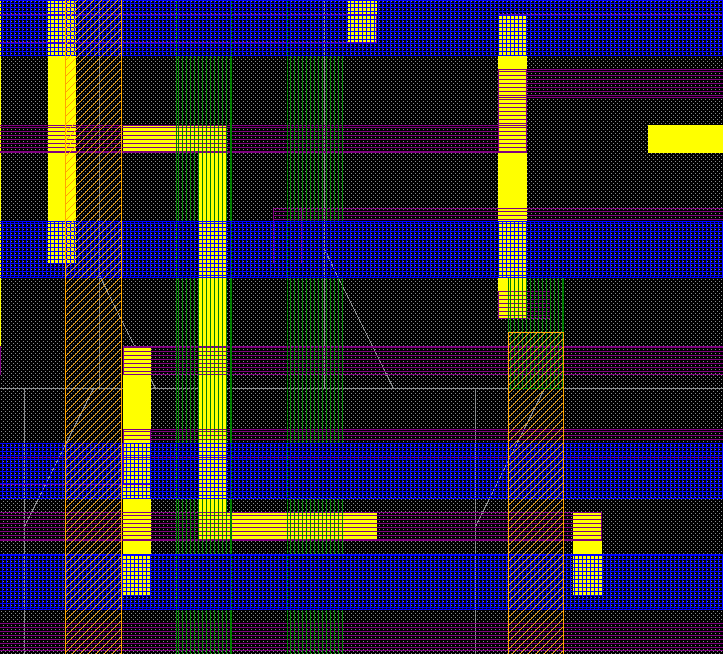}};
    \spy on (-.82,0.1) in node [left] at (1.7,-1);
    \end{tikzpicture} 
    \vspace{-2mm}
    \caption{An example of covering a net asset by flipping the cell: The exposed area (solid yellow regions in the left image) is totally covered by the nets in the upper metal layer(s) after the net is re-routed (right image).}
    \label{fig:detour}
\end{figure}

In some cases, the exposed area of the net assets can be significantly reduced by changing the orientation of the cell (flip over the Y axis). In this case, the physical synthesis tool automatically re-routes the nets connected to the flipped cell, and the chance of hiding the net asset under other nets increases, as depicted in Fig.~\ref{fig:detour}. It should be noted that this technique is performed in the very last steps of our methodology, and only the net assets with the worst exposed area are selected.

\subsection{Countermeasures against HT Insertion}

We now explain the techniques used for preventing HT insertion. Recall the \emph{exploitable region} notion, a set of continuous gaps, filler cells, unconnected cells, or non-functional cells, that an adversary can use to place his/her malicious logic. In principle, available routing resources are also considered when determining exploitable regions since the HT circuitry needs to be somehow connected to the original design. However, our efforts are directed toward mitigating the free placement sites. This is mainly because if we eliminate all the gaps, there would be no room for the HT cells to be placed in the design. Therefore, available routing resources become meaningless. 

\setcounter{subsubsection}{6}
\subsubsection{Location-based Buffering} 

Even after shrinking the design area to make the core area as dense and compact as possible, several gaps may still exist and create large exploitable areas. As the threshold for continuous gaps to be considered an exploitable region is 20 placement sites, we developed a script to search for these regions and insert buffers to fill or cut the continuous gaps below 20 sites. It should be noted that whenever buffer insertion is considered, there are overheads in power and, potentially, in timing. However, the trade-off between security and PPA is arguably beneficial for this particular technique. 

\subsubsection{Final Cell Refinement} There might be cases where the buffer insertion fails due to a lack of routing resources in congested areas. It may also succeed but create timing violations. We try to eliminate the remaining exploitable regions in these cases by slightly moving the surrounding cells. This simple technique can be done using algorithmic approaches as introduced in \cite{ASSURER} or manually by the physical design engineer (if there are few cases). 

\begin{figure*}[ht]
    \centering
   \subfloat[]{\begin{tikzpicture}[spy using outlines={circle,black!30,magnification=4,size=3cm, connect spies}]
    \node {{\includegraphics[ height=.60\columnwidth]{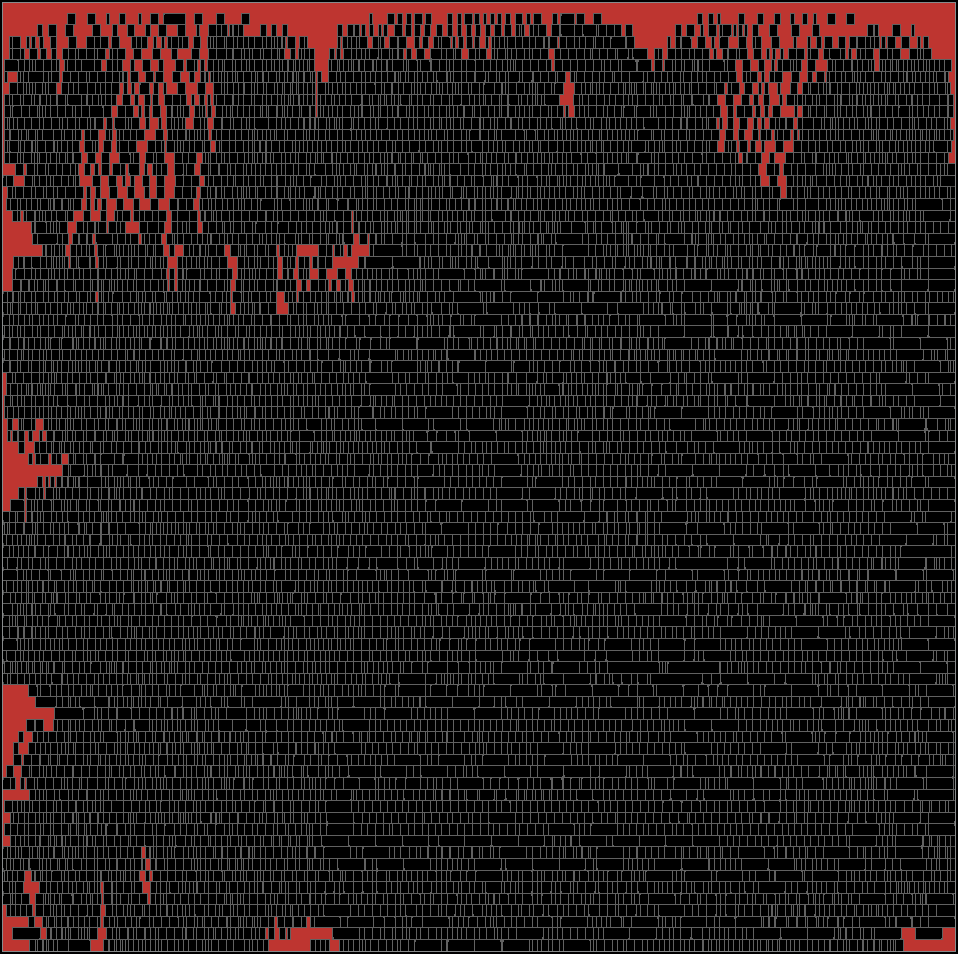}}};
    \spy on (1.6,2) in node [left] at (4.0,-1);
    \end{tikzpicture}}
    \subfloat[]{\begin{tikzpicture}[spy using outlines={circle,black!30,magnification=4,size=3cm, connect spies}]
    \node {{\includegraphics[ height=.60\columnwidth]{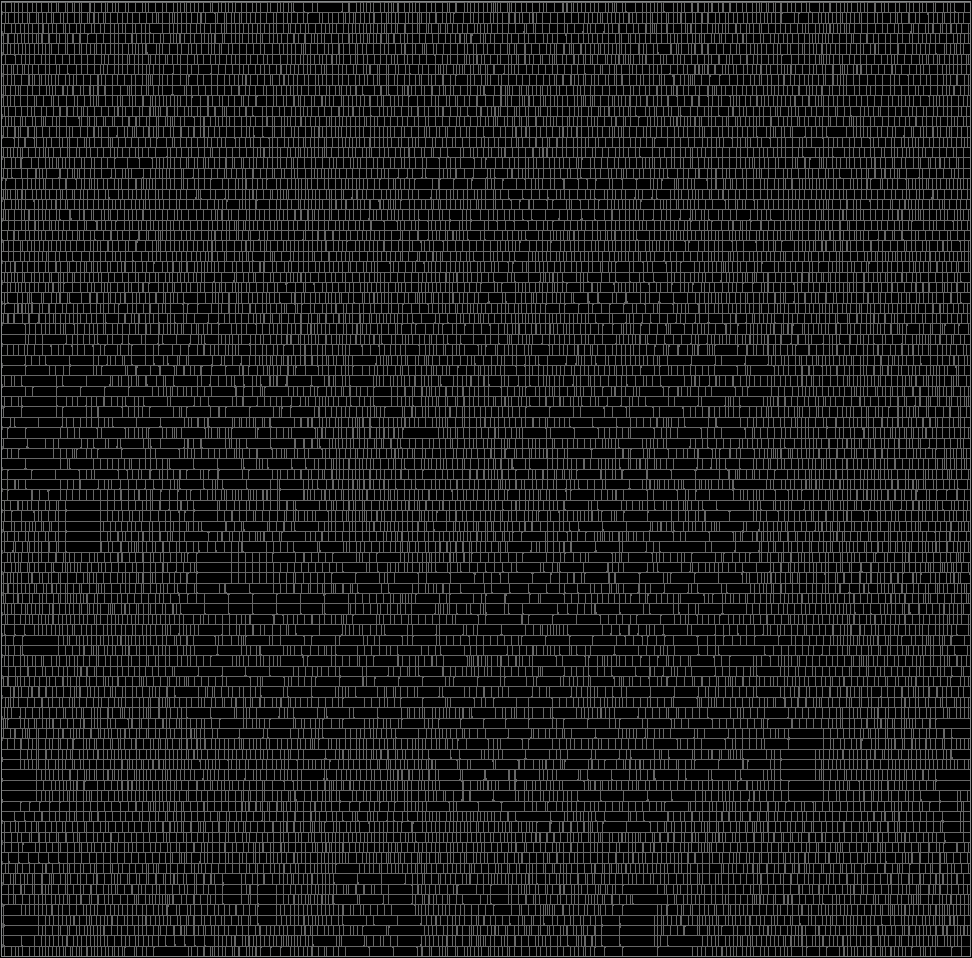}}};
    \spy on (1.6,2) in node [left] at (-1,1);
    \end{tikzpicture}} 
    \vspace{-1mm}
    \caption{An example of a) design with exploitable regions (highlighted in red), and b) design with zero exploitable regions using our techniques.}
    \label{fig:zero}
\end{figure*}

Fig.~\ref{fig:zero} shows how we eliminate all exploitable regions in a design by adopting the mentioned techniques.

\subsection{ISPD Contest Wrap-up}

The scoring formula, given in Eq.~\ref{EQ1}, was meant to normalize results against the baselines while giving equal importance to design quality and security. However, since there is a multiplicative factor between these two score components, achieving a score of 0 in security (which is impossible in practice) would also bring the overall score to 0.

Contestants eventually realized that drawing a single metal plate above the entire design was sufficient to eliminate all threats. This solution corresponds to using an entire metal layer as a sacrificial layer. This solution does create DRC violations, but the scoring formula fails to penalize teams: since the security score is zero, the design component of Eq.~\ref{EQ1} is irrelevant. All in all, the contest ended with four teams tied with ``perfect scores'' of zero. The final scores are shown in Tab.~\ref{tab:ispd_scores}. Our team is identified as team `K'.

To be very clear, we emphasize that the sacrificial metal layer solution has no practical or academic merit. It does not effectively protect against any of the considered threats. It is only effective in satisfying the contest's scoring. A clearer picture of the overheads imposed by our techniques can be seen in Tab.~\ref{tab:ispd_scores_des}, where it can be seen that when considering only the design component of Eq.~\ref{EQ1}, our scores are rather competitive. These scores were later used as a tiebreaker to define the ranking of the tied teams.

%By taking advantage of this bug in the formula, some teams only aimed for achieving a 0 score in security, regardless of how much overheads they might impose on the design quality. To this aim, they should 1) eliminate all the exploitable regions, and 2) cover all the net and cell assets (100\% protection of exposed areas) in all designs. While challenging, the first can be done using some techniques as we explained. But the latter one, with regards to the high number of sensitive elements (cell and net assets) in each design, is mostly impossible to reach in practice. However, some teams reached a solution in which they dedicate a large piece of metal in the top layer to a non-asset net, and this large wire covers the whole area of the design. Hence, four teams could get the perfect 0 overall score for all benchmarks (Table \ref{ispd_scores}). 

%While using the mentioned trick for covering the cell and net assets does not directly violate any of the contest rules, it violates the spirit of the contest which is enhancing the security as well as maintaining the quality of the design. However, there should be some constraints in the library that avoid such implementations, but since Nangate library considers the large piece of wire valid, the organizers decided to accept such submissions. To have a better understanding of the effectiveness of the techniques that the teams used, the individual scores for the design quality of the submissions are shown in Table \ref{tab:ispd_scores_des}. T

\begin{table}[ht]
\caption{Overall scores of the participating teams}
%\vspace{-2mm}
\label{tab:ispd_scores}
\centering
\begin{adjustbox}{width=0.47\textwidth}
\begin{tabular}{l|c|c|c|c|c|c|c|c}
\rule{0pt}{10pt} Benchmarks / Teams & J       & N     & O     & E     & L     & A     & Q     & K     \\ \hline
\rowcolor[HTML]{EFEFEF} \rule{0pt}{8pt} AES\_1           & 0.764 & 0.025 & 0.000 & 0.000 & 0.271 & 0.000 & 0.000 & 0.000 \\
\rule{0pt}{8pt} AES\_2           & 1.687 & 0.054 & 0.000 & 0.000 & 0.324 & 0.000 & 0.000 & 0.000 \\
\rowcolor[HTML]{EFEFEF} \rule{0pt}{8pt} AES\_3           & 1.332 & 0.000 & 0.000 & 0.000 & 0.295 & 0.000 & 0.000 & 0.000 \\
\rule{0pt}{8pt} Camellia         & 0.676 & 0.000 & 0.000 & 0.000 & 0.281 & 0.000 & 0.000 & 0.000 \\
\rowcolor[HTML]{EFEFEF} \rule{0pt}{8pt} CAST             & 1.687 & 0.000 & 0.000 & 0.000 & 0.300 & 0.000 & 0.000 & 0.000 \\
\rule{0pt}{8pt} MISTY            & 3.178 & 0.000 & 0.000 & 0.000 & 0.254 & 0.000 & 0.000 & 0.000 \\
\rowcolor[HTML]{EFEFEF} \rule{0pt}{8pt} OpenMSP430\_1    & 0.841 & 0.000 & 0.000 & 0.000 & 0.344 & 0.000 & 0.000 & 0.000 \\
\rule{0pt}{8pt} PRESENT          & 0.629 & 0.000 & 0.000 & 0.000 & 0.319 & 0.000 & 0.000 & 0.000 \\
\rowcolor[HTML]{EFEFEF} \rule{0pt}{8pt} SEED             & 2.203 & 0.000 & 0.000 & 0.000 & 0.207 & 0.000 & 0.000 & 0.000 \\
\rule{0pt}{8pt} TDEA             & 0.596 & 0.003 & 0.000 & 0.000 & 0.246 & 0.000 & 0.002 & 0.000 \\
\rowcolor[HTML]{EFEFEF} \rule{0pt}{8pt} OpenMSP430\_2    & 1.031 & 0.000 & 0.000 & 0.000 & 0.822 & 0.000 & 0.000 & 0.000 \\
\rule{0pt}{8pt} SPARX            & 0.476 & 0.000 & 0.000 & 0.000 & 0.262 & 0.000 & 0.000 & 0.000
\end{tabular}
\end{adjustbox}
\end{table}

\begin{table}[ht]
\caption{Design quality scores of the participating teams}
%\vspace{-2mm}
\label{tab:ispd_scores_des}
\centering
\begin{adjustbox}{width=0.47\textwidth}

\begin{tabular}{l|c|c|c|c|c|c|c|c}
\rule{0pt}{10pt} Benchmarks / Teams & J       & N     & O     & E     & L     & A     & Q     & K     \\ \hline
\rowcolor[HTML]{EFEFEF}  \rule{0pt}{8pt} AES\_1           & 0.995 & 0.713 & 0.447 & 0.475 & 0.527 & 0.519 & 1.347 & 0.481 \\
\rule{0pt}{8pt} AES\_2           & 3.737 & 0.702 & 0.425 & 0.458 & 0.539 & 0.509 & 0.817 & 0.461 \\
\rowcolor[HTML]{EFEFEF}  \rule{0pt}{8pt} AES\_3           & 2.689 & 1.059 & 0.473 & 0.498 & 0.566 & 0.541 & 1.171 & 0.523 \\
\rule{0pt}{8pt} Camellia         & 0.753 & 0.746 & 0.398 & 0.420 & 0.470 & 0.418 & 0.960 & 0.530 \\
\rowcolor[HTML]{EFEFEF}  \rule{0pt}{8pt} CAST             & 1.663 & 0.851 & 0.412 & 0.409 & 0.463 & 0.439 & 0.908 & 0.495 \\
\rule{0pt}{8pt} MISTY            & 5.009 & 0.753 & 0.418 & 0.396 & 0.457 & 0.417 & 1.559 & 0.458 \\
\rowcolor[HTML]{EFEFEF}  \rule{0pt}{8pt} OpenMSP430\_1    & 0.756 & 0.656 & 0.406 & 0.440 & 0.490 & 0.469 & 1.025 & 0.632 \\
\rule{0pt}{8pt} PRESENT          & 0.752 & 0.693 & 0.359 & 0.427 & 0.465 & 0.446 & 1.009 & 0.306 \\
\rowcolor[HTML]{EFEFEF}  \rule{0pt}{8pt} SEED             & 1.917 & 0.892 & 0.416 & 0.442 & 0.418 & 0.442 & 0.924 & 0.522 \\
\rule{0pt}{8pt} TDEA             & 0.750 & 0.846 & 0.459 & 0.526 & 0.534 & 0.524 & 0.808 & 0.584 \\
\rowcolor[HTML]{EFEFEF}  \rule{0pt}{8pt} OpenMSP430\_2    & 0.995 & 0.777 & 0.464 & 0.543 & 0.524 & 0.570 & 0.848 & 0.608 \\
\rule{0pt}{8pt} SPARX            & 0.753 & 0.663 & 0.397 & 0.420 & 0.422 & 0.404 & 1.047 & 0.509
\end{tabular}
\end{adjustbox}
\end{table}

\section{Silicon Validation of SALSy} 
\label{sec4}

In the previous section, we provided details about the different techniques that can be used to improve the security of an IC based on specific evaluation metrics \cite{contest}. However, all these techniques were initially developed for open-source PDKs. Commercial PDKs that are used by industry are more detailed than the academic ones and using these commercial PDKs increases the design complexity and introduces certain practical limitations. %and the practical use of them in state-of-the-art cases is questionable. 
Hence, we decided to fabricate a chip with the mentioned security features to highlight the gaps and limitations with the open PDKs and provide solutions to cope with these limitations.  

\begin{figure}[tb]
    \centering
    \includegraphics[width=.47\textwidth, trim=0.5cm 32pt 1cm 20pt, clip]{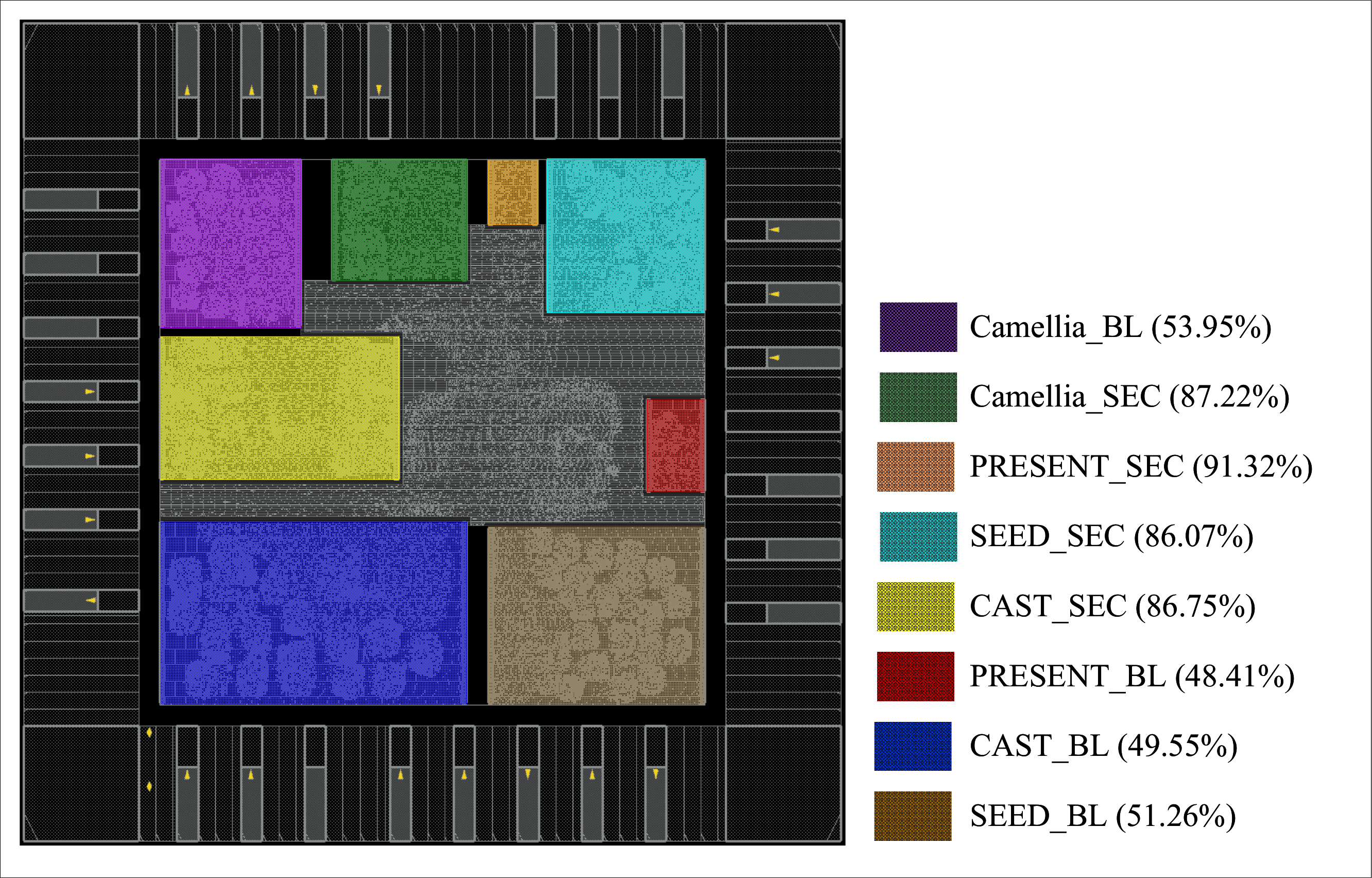}
     \vspace{-2mm}
    \caption{Floorplan view of the chip including eight blocks and density of each block.}
    \label{fig:fp}
\end{figure}

The first step in designing our chip is to localize the scoring system with respect to the commercial library so that we can evaluate the security features of the chip while adopting the same metrics from \cite{contest}. Next, we have to decide which designs we are going to tapeout. We have opted for a small chip size (1 $mm^2$) that is enough to fit four designs arranged as eight blocks: four secured versions (SEC) and four baseline versions (BL). By having a pair of each benchmark on the same chip, we can evaluate and compare the security and design quality of each block fairly. 
The designs were selected to provide variety in terms of complexity and size; The final candidates include Camellia, CAST, PRESENT, and SEED. The floorplan of the chip is shown in Fig.~\ref{fig:fp}. Different blocks are highlighted in colors and the rest of the core area is dedicated to the comparison and control unit. To make the comparison fair, we set the same density for all BL versions of the benchmarks as they were set in the contest. For the SEC variants, density is a function of the SALSy methodology. Furthermore, we implement distinct power domains for each block. This strategic approach enables us to activate only one block at a time while ensuring all other blocks remain in an off-state. Consequently, we can precisely measure the power consumption, one block at a time.  

\begin{figure}[tb]
    \centering
    \includegraphics[width=0.595\columnwidth]{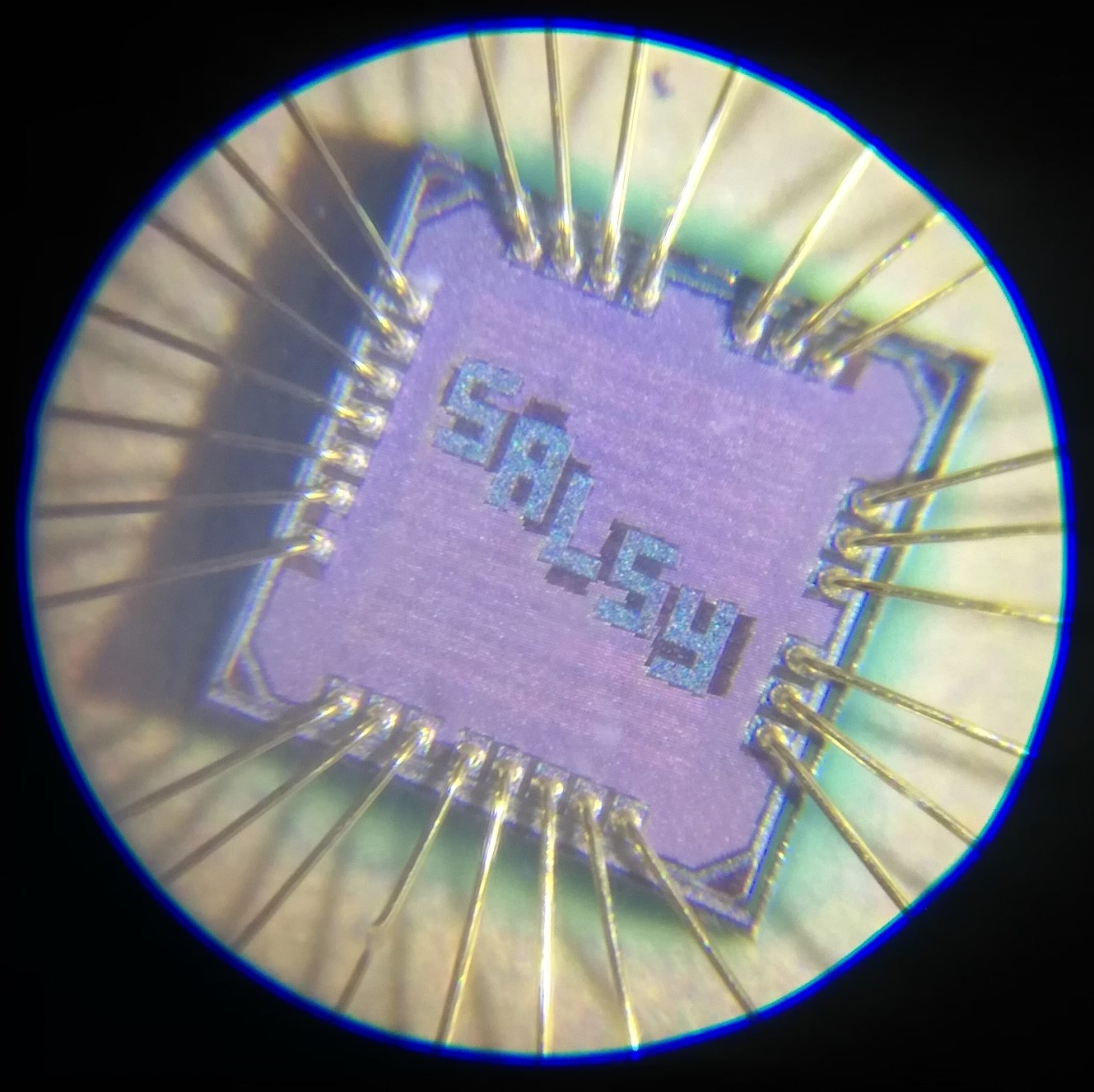}
     \vspace{-2mm}
    \caption{Microscope view of the fabricated 1mm$^2$ chip. }
    \label{chip}
\end{figure}

A microscope view of our fabricated chip is shown in Fig.~\ref{chip}. We validate the efficiency of our techniques on all four benchmarks implemented on our chip. However, it should be noted again that our approach is applicable to any design, regardless of  function or size. Another advantage of our solution is that it does not need previous knowledge about the design since it is performed at the layout level. We hypothesize that SALSy creates an opportunity to assign security closure to a separate design team since no specific details/characteristics of the design are needed for improving security. The interface between this team and the traditional physical synthesis team would be the list of assets. 

\subsection{Implementation}

A key factor to improve HT and FSP/FI security is to increase the design's density. By doing this, first, the chance of HT insertion decreases by reducing the number of gaps, and second, more cell and net assets can be protected against FI/FSP due to the increased wire congestion. Therefore, we shrink all designs as much as possible before applying any specific technique. Note the smaller size of the SEC variants with respect to the BL variants in Fig.~\ref{fig:fp}. In the following, we provide more details about the chip implementation. It should be noted that the related scripts for the implementation flow have become publicly available in \cite{github}.

\subsubsection{Non-default Rule CTS}

This technique is applicable to commercial libraries as well but with some restrictions regarding the maximum width of the wires. These restrictions are imposed by the foundry to maintain design compatibility with their equipment. In other words, we cannot enlarge the clock wires by arbitrary factors as we did in the contest, but still, it is possible to enlarge them from their default size. Although this technique is less effective in compared with the contest, we decided to keep using it since it has negligible overhead on the power and other performance-related metrics.

\subsubsection{Layer-targeted Routing}

Similar to what we did in the contest, we use the routing strategy from Algorithm \ref{alg1}. The only difference here, when using commercial libraries, is that they are more detailed and characterized to ensure the design rule accuracy and verification. Hence, as the design gets denser, more violations appear due to different reasons to ensure the quality and reliability of the chip during the fabrication process. This means that achieving high density (above 90\% for the considered 65nm technology) is very challenging. As a consequence, it is impossible to route all the asset or non-asset nets in their preferred metal layers as we did in the contest. However, even with the cost of routing some of the asset nets in the top metal layers, this technique still helps to cover a large portion of the exposed area of the assets (see Section \ref{sec5} for more details).

\subsubsection{Multicut Via Insertion}

This method is the first one we had to abandon due to the strict constraint in the commercial library. Although it is theoretically possible to use multi-cut vias to connect the pins, it creates DRC issues after the wires are connected to the vias. Given the significant challenge of addressing numerous DRC violations, adopting this method in our chip was not practical. Yet, this solution can be revisited for a different commercial library/PDK.

%One solution to cope with these DRC violations is to re-route the problematic wires one by one, which needs a huge amount of time and effort since it cannot be automated. Hence, we gave up using this method in our chip implementation.

\subsubsection{Edge Cell Placement}

An important difference between the contest and an actual tapeout is that each design was treated as a separate chip in the contest, while we have to put all the designs together on one chip. Hence, we have little freedom in defining the IO pin locations for each design. The decision of putting the IO pins on one of the block sides (left, right, bottom, or top) is defined during top-level floorplanning. For instance, if we consider the location of the \emph{PRESENT\_SEC}  block in Fig. \ref{fig:fp}, the best place to put the IO pins is on the bottom edge of the block since it has the closest distance to the control and comparison unit which is located in the center of the chip. In this case, the routing would be done with fewer issues and unnecessary resource utilization would be avoided, leading to a more optimized floorplan. 

\begin{figure}[tb]
    \centering
    \includegraphics[height= .64\columnwidth, trim=0 0 0 0.5cm, clip]{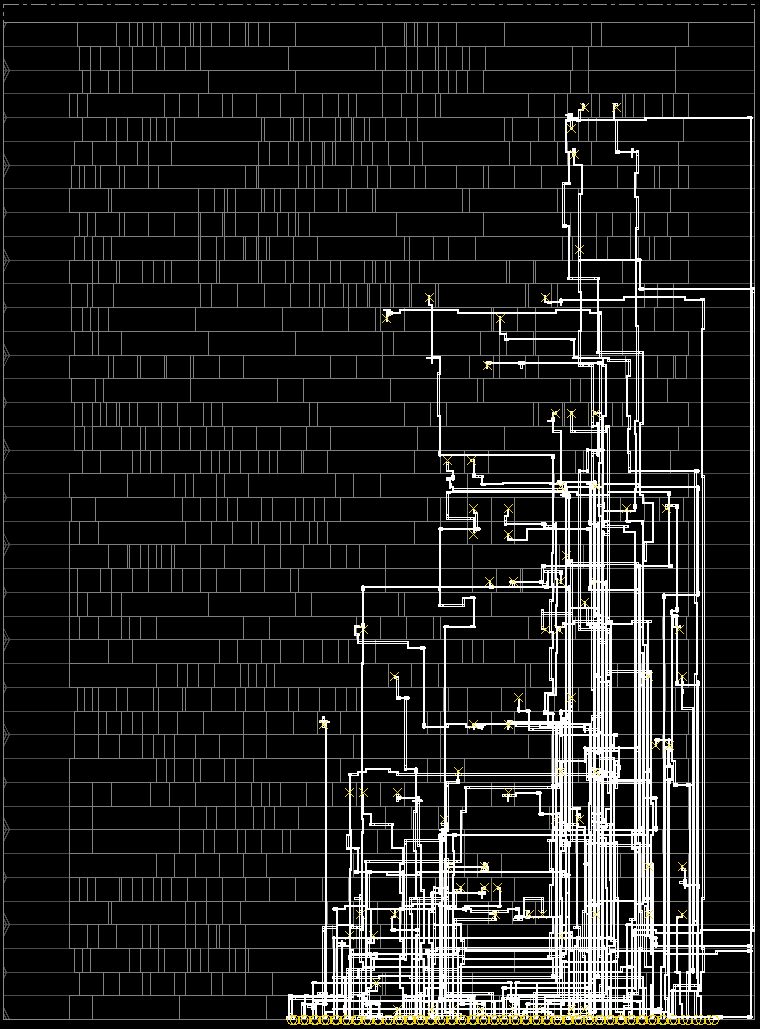}
    \vspace{-2mm}
    \caption{A design (PRESENT) with most of the IO pins on the bottom side and the net assets (highlighted in white) connected to their relative IO pin.}
    \label{present_chip}
\end{figure}

This restriction in pin placement leads to the limitation of using this technique in our chip. As shown in Fig. \ref{present_chip}, there is only a limited area close to the IO pins that are connected to the net assets (highlighted in white), and it is impossible to fit all the connected cells near their driver/sink pins since the congestion in that region increases significantly and the design becomes unroutable. Hence, this technique cannot be used in this specific floorplan. However, it is crucial to highlight that this limitation does not impede the potential application of this technique in other chip implementations, as in the contest that we could fit the cells near to their relative IO pins due to the square shape of the block and the fact that there was enough space around all four sides of the design (Fig. \ref{fig:4techs}). 

\subsubsection{Selectively Cell Flipping}

This technique can be used in chip implementation as well as in the contest setting without facing any specific limitations. However, we minimize the use of this method due to its inherent manual nature. Our overarching objective is to uphold a holistic and automated approach, eschewing the adoption of selective techniques, in order to ensure the comprehensive applicability of our approach.
%because it is mostly a manual effort and we want to keep our approach in a general and automated fashion and avoid selective methods. 

\subsubsection{Intermediate Buffering}

As mentioned in the previous section, buffer insertion can have undesirable effects on the timing and power of the design. In the contest, such issues were only considered as a negative factor in the final score to penalize the teams. But in the actual chip, any single issue that violates the timing of the design (i.e. setup time, hold time, etc.) was considered unacceptable. Hence, the timing closure of the design should be perfect and the trade-off between the timing issues and the enhanced security is only possible as long as the timing slack remains positive. Due to this reason, we replaced this technique with a smarter buffer insertion algorithm which is explained in the text that follows.

\subsubsection{Location-based Buffering}

\begin{figure}[tbp]
\centering
\begin{tikzpicture}[spy using outlines={circle,black!30,magnification=4,size=3cm, connect spies}]
\node {\includegraphics[width=0.335\textwidth, keepaspectratio]{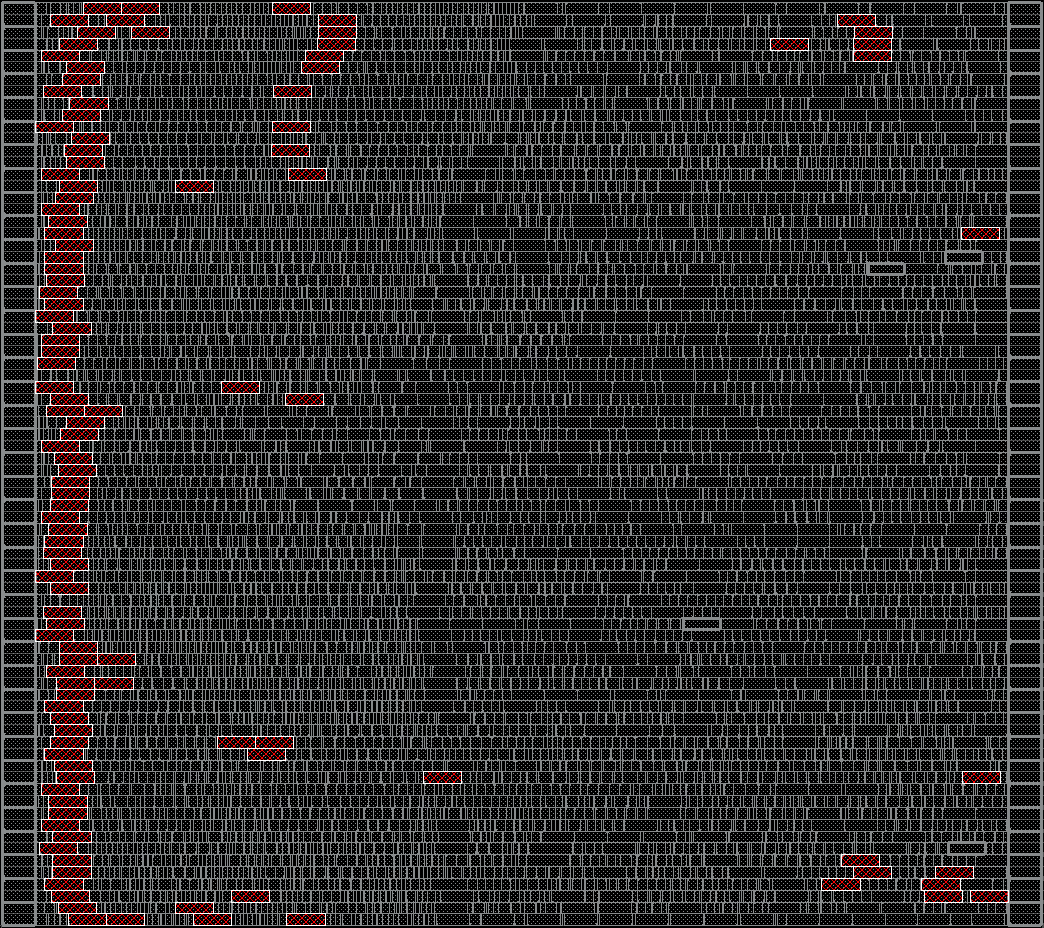}};
\spy on (-2.6,-2.35) in node [left] at (2,1);
\end{tikzpicture}    
 \vspace{-2mm}
\caption{Added buffers (highlighted in red) using our smart algorithm to eliminate exploitable regions.}
    \label{buf_chip}
\end{figure}

As mentioned in the previous part, we change our buffer insertion strategy such that it is no longer aimed to shorten the long net assets. Instead, it is totally focused on filling the continuous gaps of the design in a completely automated fashion. In other words, the buffer insertion technique turns into a location-based algorithm that seeks exploitable regions, rather than searching for  long net assets. The sinks and the drivers of the added buffer are selected from the nearby cells such that it has the least negative impact on the timing of the design, as highlighted in Fig.~\ref{buf_chip}. 

\subsubsection{Final Cell Refinement}

Similar to what we did in the contest, this technique is used in the very late steps of the chip implementation as manual fixes. If any exploitable region is left that can be eliminated with a few cell movements, it is possible to use this method. However, we decided to minimize this technique in our chip due to two reasons: i) if an exploitable region is eliminated simply by moving the cells, the adversary can revert the changes to make enough space for his/her malicious logic as well. Although being effective in the contest, in a realistic scenario it seems to be a useless effort; ii) It  conflicts with our aim to create an automated flow.

\section{Results} \label{sec5}

In this section, we provide the experimental results obtained from  chip design and measurements. During the physical implementation process, we used Cadence and Siemens toolchains and our target technology is a general-purpose flavor of a 65nm CMOS technology.

We divide the results into two parts, pre-silicon and post-silicon results. The first part represents the results obtained from the final layout sent for fabrication such as the area and density of the  blocks. Physical chip measurements, such as power consumption, are provided in the second part.

\subsection{Pre-silicon Results}

As mentioned before, we evaluate our techniques using the same scoring system used in \cite{contest}. In the following, we provide more details about each metric. 

Based on the considered metrics, the final scores of our approach are presented in Tab.~\ref{table3}. The table clearly demonstrates that our best scores are related to HT insertion, providing strong evidence that SALSy can effectively function as a prevention technique in a realistic PDK setting. Conversely, as anticipated, our worst scores are associated with power, primarily resulting from the buffer insertion applied to enhance security. The power consumption of our designs consistently exceeded the baselines, as indicated by values greater than 1.0 for all cases. Nevertheless, it is important to acknowledge that a certain level of overhead is inevitable when trading off for enhanced security. The power numbers reported in Tab.~\ref{table3} are estimates from physical synthesis; precise power consumption numbers are provided in Section~\ref{subsec:measurements}.

%MISSING
The presented table illustrates a substantial reduction in the count of \emph{exploitable regions} within our secured version across all benchmarks. This reduction is notably profound, with a 100\% decrease observed in the case of Camellia and PRESENT benchmarks. Furthermore, for the CAST and SEED benchmarks, the reduction percentages stand at 95.3\% and 90.3\% respectively.
Regarding FSP/FI assessment, benchmarks exhibit varying outcomes. PRESENT benchmark excels with an impressive 43\% \emph{exposed area} reduction from the baseline, while the CAST benchmark demonstrates a more moderate 18.5\% enhancement. 

\begin{table*}[ht]
\caption{Final scores of our approach for four different benchmarks}
 \vspace{-2mm}
\label{table3}
\centering
\begin{tabular}{lllcccc}

\hline
\multicolumn{3}{c}{\rule{0pt}{8 pt}Metrics / Benchmarks}     & \multicolumn{1}{l}{Camellia}  & \multicolumn{1}{l}{CAST}  & \multicolumn{1}{l}{PRESENT} & \multicolumn{1}{l}{SEED} \\ \hline 

\multicolumn{1}{l}{\multirow{5}{*}{Design Quality}}  & \multicolumn{1}{l} {\rule{0pt}{8pt}DRC}  & des\_issues   & \cellcolor[HTML]{C6E4BD}0.000     & \cellcolor[HTML]{C6E4BD}0.000     & \cellcolor[HTML]{C6E4BD}0.000     & \cellcolor[HTML]{C6E4BD}0.000 \\ \cline{2-7} 

\multicolumn{1}{l}{}    & \multicolumn{1}{l}{\multirow{3}{*}{PPA}}  & \rule{0pt}{8pt}des\_perf      & \cellcolor[HTML]{C6E4BD}0.000     & \cellcolor[HTML]{C6E4BD}0.000     & \cellcolor[HTML]{C6E4BD}0.000     & \cellcolor[HTML]{C6E4BD}0.000 \\

\multicolumn{1}{l}{}    & \multicolumn{1}{l}{}  & \rule{0pt}{8pt}des\_p\_total  &  \cellcolor[HTML]{FFCCC9}1.184     &  \cellcolor[HTML]{FFCCC9}1.072     &  \cellcolor[HTML]{FFCCC9}1.161     &  \cellcolor[HTML]{FFCCC9}1.041 \\

\multicolumn{1}{l}{}    & \multicolumn{1}{l}{}  & \rule{0pt}{8pt}des\_area      & \cellcolor[HTML]{f7e5af}0.686     & \cellcolor[HTML]{f7e5af}0.606     & \cellcolor[HTML]{f7e5af}0.597     & \cellcolor[HTML]{f7e5af}0.627 \\ \cline{2-7} 

\multicolumn{1}{l}{}    & \multicolumn{1}{l}{\rule{0pt}{8pt}Overall}    & des   & \cellcolor[HTML]{ebf7af}0.467     & \cellcolor[HTML]{ebf7af}0.419     & \cellcolor[HTML]{ebf7af}0.439     & \cellcolor[HTML]{ebf7af}0.417 \\ \hline

\multicolumn{1}{l}{\multirow{6}{*}{Security}}       & \multicolumn{1}{l}{\multirow{2}{*}{Trojan Insertion}}     & \rule{0pt}{8pt}ti\_sts    & \cellcolor[HTML]{C6E4BD}0.000     & \cellcolor[HTML]{C6E4BD}0.015     & \cellcolor[HTML]{C6E4BD}0.000     & \cellcolor[HTML]{C6E4BD}0.026 \\

\multicolumn{1}{l}{}    & \multicolumn{1}{l}{}  & \rule{0pt}{8pt}ti\_fts        & \cellcolor[HTML]{C6E4BD}0.000     & \cellcolor[HTML]{C6E4BD}0.079     & \cellcolor[HTML]{C6E4BD}0.000     & \cellcolor[HTML]{e0f1db}0.169 \\ \cline{2-7}

\multicolumn{1}{l}{}    & \multicolumn{1}{l}{\rule{0pt}{8pt}Overall}  & \rule{0pt}{8pt}ti             & \cellcolor[HTML]{C6E4BD}0.000     & \cellcolor[HTML]{C6E4BD}0.047    & \cellcolor[HTML]{C6E4BD}0.000     & \cellcolor[HTML]{C6E4BD}0.097 \\ \cline{2-7}

\multicolumn{1}{l}{}    & \multicolumn{1}{l}{\multirow{2}{*}{FSP/FI}} & \rule{0pt}{8pt}fsp\_fi\_ea\_c & \cellcolor[HTML]{f8da7e}0.842     & \cellcolor[HTML]{f8da7e}0.797     & \cellcolor[HTML]{e0f1db}0.293     & \cellcolor[HTML]{f8da7e}0.762 \\

\multicolumn{1}{l}{}    & \multicolumn{1}{l}{}  & \rule{0pt}{8pt}fsp\_fi\_ea\_n & \cellcolor[HTML]{f7e5af}0.624     &  \cellcolor[HTML]{f8da7e}0.833     & \cellcolor[HTML]{f7e5af}0.568     & \cellcolor[HTML]{f8da7e}0.835 \\ \cline{2-7} 

\multicolumn{1}{l}{}    & \multicolumn{1}{l}{\rule{0pt}{8pt}Overall}    & fsp\_fi        & \cellcolor[HTML]{f8da7e}0.733    & \cellcolor[HTML]{f8da7e}0.815     & \cellcolor[HTML]{ebf7af}0.430     & \cellcolor[HTML]{f8da7e}0.799 \\ \hline

\multicolumn{2}{l}{Final score}     & \rule{0pt}{8pt}OVERALL        & \cellcolor[HTML]{e0f1db}0.171     & \cellcolor[HTML]{e0f1db}0.181     & \cellcolor[HTML]{C6E4BD}0.094     & \cellcolor[HTML]{e0f1db}0.187    \\
\hline
\end{tabular}
% \end{adjustbox}
\end{table*}

To provide a comprehensive understanding of the relationship between each step of SALSy and the resulting scores, individual scores for the PRESENT benchmark are presented in Tab.~\ref{table4} after subsequently applying each technique. The table reveals that the Layer-targeted Routing step has the most significant effect on the \emph{fsp\_fi} and overall scores, given its substantial impact on increasing congestion. On the other hand, the Location-based Buffer Insertion technique has the most substantial impact on enhancing the \emph{ti} score, as it drastically reduces the number of gaps in the layout. Remarkably, the overall trend of score improvement, as displayed in Table \ref{table4}, remains consistent for all other three benchmarks. We omit these results for the sake of space.

\begin{table*}[ht]
\caption{The changes in the scores of PRESENT benchmark after applying our techniques in each step}
 \vspace{-2mm}
\label{table4}
\centering
\begin{tabular}{lllcccc}

\hline
\multicolumn{3}{c}{\rule{0pt}{8pt}Metrics / Steps}     & \multicolumn{1}{c}{\begin{tabular}[c]{@{}c@{}}Non-default Rule \\ CTS\end{tabular}}  &  \multicolumn{1}{c}{\begin{tabular}[c]{@{}c@{}}Layer-targeted\\  Routing\end{tabular}}  & \multicolumn{1}{c}{\begin{tabular}[c]{@{}c@{}}Location-based \\ Buffer Insertion\end{tabular}} & \multicolumn{1}{c}{\begin{tabular}[c]{@{}c@{}}Final  \\ Cell Refinement\end{tabular}} \\ 
\hline 

\multicolumn{1}{l}{\multirow{5}{*}{Design Quality}}  & \multicolumn{1}{l} {\rule{0pt}{8pt}DRC}  & des\_issues   &    0.000     &    0.000     &    0.000     &    0.000 \\ \cline{2-7} 

\multicolumn{1}{l}{}    & \multicolumn{1}{l}{\multirow{3}{*}{PPA}}  & \rule{0pt}{8pt}des\_perf      &    0.000     &    0.000     &    0.000     &    0.000 \\

\multicolumn{1}{l}{}    & \multicolumn{1}{l}{}  & \rule{0pt}{8pt}des\_p\_total  &     1.018     &     1.138     &     1.159     &     1.161 \\

\multicolumn{1}{l}{}    & \multicolumn{1}{l}{}  & \rule{0pt}{8pt}des\_area      &    0.597     &    0.597     &    0.597     &    0.597 \\ \cline{2-7} 

\multicolumn{1}{l}{}    & \multicolumn{1}{l}{\rule{0pt}{8pt}Overall}    & des   &    0.404     &    0.434     &    0.436     &    0.439 \\ \hline

\multicolumn{1}{l}{\multirow{6}{*}{Security}}       & \multicolumn{1}{l}{\multirow{2}{*}{Trojan Insertion}}     & \rule{0pt}{8pt}ti\_sts    &    0.010     &    0.011     &    0.005     &    0.000 \\

\multicolumn{1}{l}{}    & \multicolumn{1}{l}{}  & \rule{0pt}{8pt}ti\_fts        &    0.116     &    0.117     &    0.071     &    0.000 \\  \cline{2-7}

\multicolumn{1}{l}{}    & \multicolumn{1}{l}{\rule{0pt}{8pt}Overall}  & \rule{0pt}{8pt}ti             &    0.063     &    0.064    &    0.038     &    0.000 \\ \cline{2-7}

\multicolumn{1}{l}{}    & \multicolumn{1}{l}{\multirow{2}{*}{FSP/FI}} & \rule{0pt}{8pt}fsp\_fi\_ea\_c &    0.913     &    0.318     &    0.315     &    0.298 \\

\multicolumn{1}{l}{}    & \multicolumn{1}{c}{}  & \rule{0pt}{8pt}fsp\_fi\_ea\_n &    0.985     &     0.586     &    0.583     &    0.568 \\ \cline{2-7} 

\multicolumn{1}{l}{}    & \multicolumn{1}{l}{\rule{0pt}{8pt}Overall}    & fsp\_fi        &    0.949    &    0.452     &    0.449     &    0.430 \\ \hline

\multicolumn{2}{l}{Final score}     & \rule{0pt}{8pt}OVERALL        &    0.204     &    0.112     &    0.106     &    0.094    \\
\hline

\end{tabular}
%\end{adjustbox}
\end{table*}

\subsection{Post-fabrication Results}
\label{subsec:measurements}

\begin{figure}[tb]
    \centering
    \includegraphics[width=0.8\columnwidth]{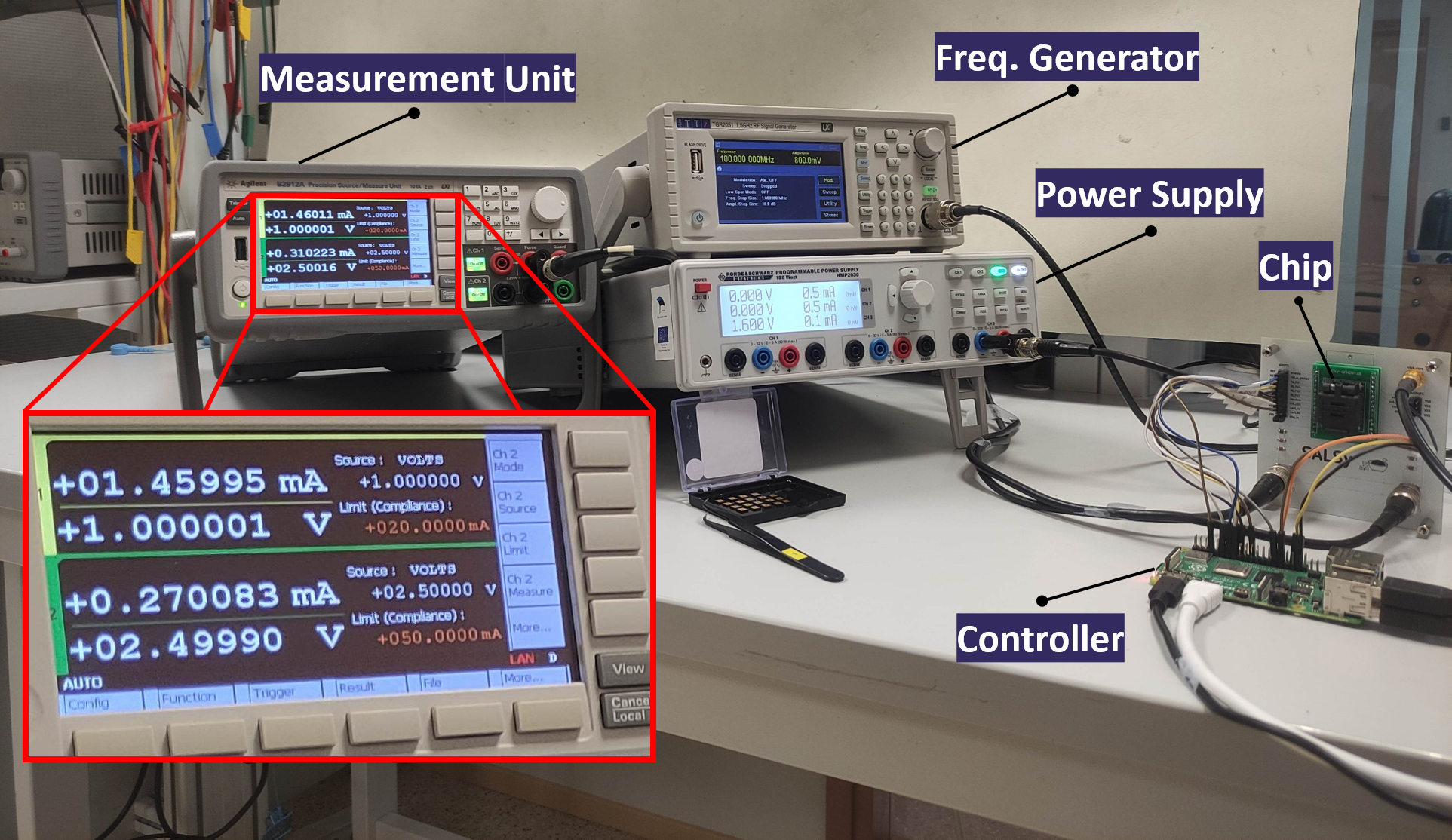}
    \vspace{-2mm}
    \caption{The testing environment for the fabricated chip.}
    \label{test}
\end{figure}

In this section, we present the measurement results obtained from the actual chip. The testing environment, illustrated in Fig. \ref{test}, comprises several components: a controller responsible for serial communication, input feeding, output reading, and data analysis; a power supply; a frequency generator providing a fast clock; and a precise measuring unit for assessing the chip's power consumption under various scenarios. We conducted the experiments on 20 packaged chips, chosen from a total of 100 fabricated chips.

\subsubsection{Verifying the Chip Functionality}

Before proceeding with power measurements, it is crucial to ensure that our chips, particularly their blocks, are functioning as intended. To accomplish this, we developed a Python script that systematically activates the blocks one by one at the target frequency, while simultaneously verifying the validity of the output data for each chosen block. All chips are deemed functional and we proceed with power measurements. It is worth mentioning that our target frequency for all blocks is 100MHz, whereas the clock frequency for the comparison and control unit is set to 1MHz. A fast 100MHz reference clock is generated by an external frequency generator, as depicted in Fig.~\ref{test}. We remind the reader that total power is the sum of dynamic and static (leakage) power, which we will report separately. Our dynamic power results are reported at 100MHz.

\subsubsection{Leakage Power Measurement}

Once the functionality of the chip has been verified, we proceed to measure the power consumption. We begin by assessing the Always On (AO) leakage power of the chip. As the name suggests, AO indicates that this type of power consumption is present consistently, regardless of whether the IC is actively performing computations or tasks (functional mode), and measuring AO leakage power does not depend on the switching activity of the transistors of the chip. In this measurement, no inputs other than the power supply signals are fed into the chip. It allows us to capture the baseline power consumption when the chip is in its idle state, and no specific operations are being performed.

%MISSING

Following the AO leakage power measurement, we proceed with measuring the leakage power of each individual block. To achieve this, we activate one block at a time by asserting the appropriate configuration of the input signals specifically designed for the voltage island of that block. Each block is encompassed by power switches, granting us the capability to activate or deactivate them as needed. %With multiple power domains integrated into the chip's design, the activated block exclusively draws power from its designated power domain. 
This meticulous power domain segregation significantly enhances measurement accuracy by eliminating power-sharing with any other block. Similar to the AO leakage measurement, no clock or any other signals are fed to the chip during this procedure. This allows us to precisely assess the power consumption of each individual block in isolation, shedding light on their specific power characteristics.

\begin{figure*}[ht]
    \centering
    \includegraphics[width=0.82\textwidth, trim=1cm 5.5cm 1cm 5.5cm, clip]{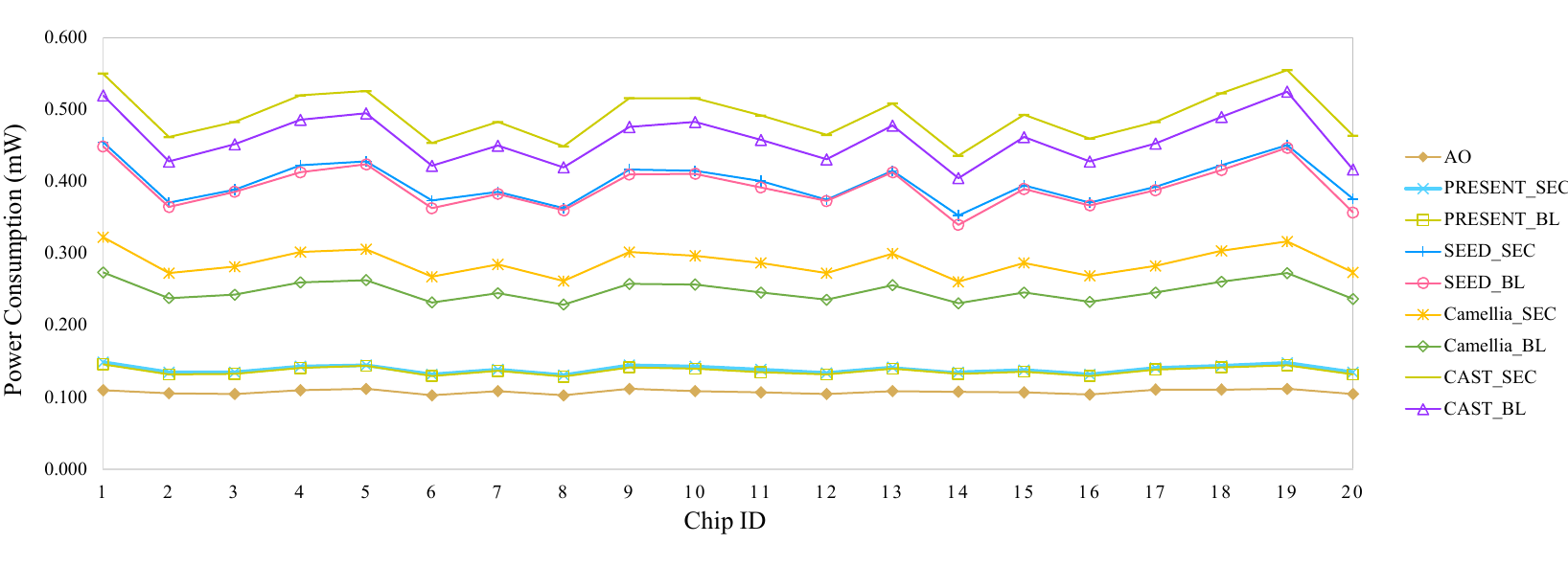}
     \vspace{-6mm}
    \caption{The measured leakage power for 20 fabricated chips (in mW).}
    \label{leakage}
\end{figure*}

The leakage power results are depicted in Fig.~\ref{leakage}. As illustrated, different chips exhibit distinct power signatures, which can be attributed to process variation. These variations are inherent in the semiconductor fabrication process and can lead to differences in power consumption among individual chips. The observed differences in leakage power highlight the significance of process variability in chip manufacturing and underscore the need for thorough testing and analysis of power characteristics in real-world chip deployments.
%MISSING
The static power incurs an average overhead of  1.72\%, 1.66\%, 15.89\%, and 7.24\% across the PRESENT, SEED, Camellia, and CAST benchmarks, respectively. 

\subsubsection{Dynamic Power Measurement}

\begin{figure*}[ht]
    \centering
    \includegraphics[width=0.82\textwidth, trim=1cm 5.5cm 1cm 5.5cm, clip]{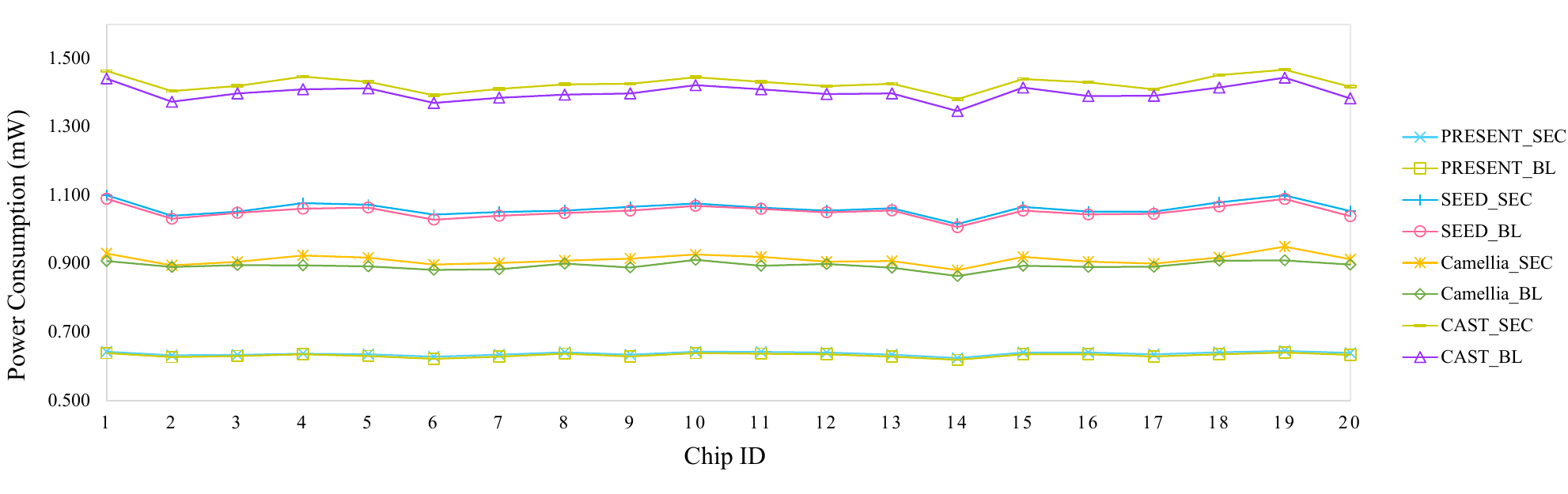}
     \vspace{-5mm}
    \caption{The measured dynamic power for 20 fabricated chips (in mW).}
    \label{dynamic}
\end{figure*}

The dynamic power measurement test is conducted to assess the power consumption of each design block in functional mode. To achieve this, we activate the blocks one by one and provide them with appropriate inputs (plain text) while operating at a clock frequency of 100MHz. The plain text inputs can be sourced either from an internal register bank within the chip or from the host controller through the UART protocol.
The results of the dynamic power measurement are presented in Fig. \ref{dynamic}.
%MISSING
Across all benchmarks, the average overhead for dynamic power consumption remains below 3\%. Specifically, the PRESENT, SEED, Camellia, and CAST benchmarks exhibit overheads of 0.79\%, 0.86\%, 2.02\%, and 1.96\% respectively.

\section{Discussion}  \label{sec6}
% For the revision: we can talk about the power measurement such that it gives realistic numbers such that we can ignore the unfair weight of power in the contest scoring

SALSy aims to prevent post-design attacks and is evaluated with the metrics from \cite{contest}. However, some introduced metrics could arguably be redefined to make the evaluation more realistic - thus, the results could be readily leveraged by industry. For instance, considering the threshold of 20 continuous gaps for the exploitable area might be too optimistic since small HTs only occupy a few placement sites \cite{A2Trojan}. Furthermore, for the FSP/FI threats, the aim is to protect the design. For instance, in the ideal case, if an attacker tries to compromise the sensitive data by drilling a hole (milling), the chip should fail to operate due to damage to the protective nets above the sensitive nets. Hence, trivial defense schemes such as covering the whole core area with a large metal plate should not be considered a valid solution since the existence of this layer will not be an obstacle for the attacker.

Moreover, the scripted nature of SALSy is a strategic choice that aligns with the scalability needs of the industry. While designing a security-aware place and route engine might be an attractive academic pursuit, such an endeavor could lack the scalability required for real-world, large-scale chip designs. By scripting SALSy, we emphasize its flexibility, making it adaptable to various design sizes and complexities, and thus addressing the practical needs of industry.

Additionally, standard deviation values for the leakage power consumption of both the baseline and secured versions of each block were obtained. The minimum values are \(5.41\times 10^{-3}\) and \(5.49\times 10^{-3}\) for the baseline and secured versions of the PRESENT benchmark, respectively, while the maximum values are \(3.46\times 10^{-2}\) for the baseline and \(3.39\times 10^{-2}\) for the secured versions of the CAST benchmark. These results highlight that the SALSy approach exhibits sensitivity to process variation comparable to that of the conventional security-unaware flow.

We have also compared SALSy against the most related prior arts to give the reader a better understanding of the key differences in power, timing, area, and density promoted by SALSy. As shown in Tab.~\ref{tableF}, our work is the only one that validated the presented technique in silicon. All other works only aim for security closure, and some of them suffer from the various issues we have highlighted related to the use of limited PDKs/libraries. 
The {\color[HTML]{009900}\Smiley} sign indicates improvement, the {\color[HTML]{fe0000}\Sadey} sign indicates deterioration of the introduced metrics, while the \Neutrey sign indicates that there are no considerable changes after applying the individual technique. N/A indicates the metric is not reported by the authors. It should be noted that we consider the increase in density as an improvement since it enhances the security of the design against the considered threats.

\begin{table*}[ht]
\label{tableF}
\caption{Comparison of this work (SALSy) with the previous techniques}
 \vspace{-3mm}
\centering
    \begin{adjustbox}{width=0.7\textwidth}

\begin{tabular}{llccccc}
Ref.    & Technique     & \multicolumn{4}{c}{Implications} & Validated? \\ \hline

{\cite{tehraniFSP}} & {Internal Shielding} & Power \color[HTML]{fe0000}\Sadey        & Timing \color[HTML]{fe0000}\Sadey            & Area \color[HTML]{fe0000}\Sadey          & Density \color[HTML]{009900}\Smiley      & {\color[HTML]{fe0000}\xmark}                     \\

{\cite{second}}    & {TroMUX}      & Power \color[HTML]{fe0000}\Sadey    & Timing \color[HTML]{fe0000}\Sadey   & Area \Neutrey           & Density \color[HTML]{009900}\Smiley          & {\color[HTML]{fe0000}\xmark} \\                          

{\cite{bisa}}   & {BISA}& Power \color[HTML]{fe0000}\Sadey     & N/A    & Area \Neutrey  & Density \color[HTML]{009900}\Smiley   & {\color[HTML]{fe0000}\xmark}\\

{\cite{papa}}         & {Layout Filling}& Power \color[HTML]{fe0000}\Sadey    & Timing \color[HTML]{fe0000}\Sadey  & Area \Neutrey  & Density \color[HTML]{009900}\Smiley  & {\color[HTML]{fe0000}\xmark}  \\

{\cite{JohannICCAD}}         & {DEFense Framework}   & Power \Neutrey    & Timing \color[HTML]{fe0000}\Sadey    & Area \Neutrey    & Density \color[HTML]{009900}\Smiley  & {\color[HTML]{fe0000}\xmark} \\

{\cite{ASSURER}}         & {ASSURER}    & Power \Neutrey      & N/A  
    & Area \color[HTML]{009900}\Smiley      & Density \color[HTML]{009900}\Smiley         & {\color[HTML]{fe0000}\xmark} \\

{\cite{ter}}    & {T-TER}      & Power \Neutrey    & Timing \Neutrey   & Area \Neutrey          & Density \Neutrey         & {\color[HTML]{fe0000}\xmark} \\

     & {{This Work (SALSy)}}    & Power \color[HTML]{fe0000}\Sadey     & Timing \color[HTML]{009900}\Smiley      & Area \color[HTML]{009900}\Smiley  & Density \color[HTML]{009900}\Smiley   & {\color[HTML]{009900}\cmark}
     
\end{tabular}
\end{adjustbox}
\end{table*}

\section{Conclusion}  \label{sec7}

In this paper, we have introduced SALSy, a design-time methodology to bolster the security of ICs against fabrication-time and post-fabrication attacks. Through a silicon demonstration, we successfully validated our solution, showcasing its fitness for use with a commercial PDK and cell library. 
In our pursuit of heightened security, our methodology strikes a prudent balance by incurring only a modest increase in power consumption. Although effective against Trojan insertion, there is still room for enhancing security against FSP/FI. Our future research will focus on automating the introduced techniques, including selective methods, and incorporating new approaches to bolster FSP/FI defense and overall security. Additionally, we contemplate introducing new evaluation metrics alongside the existing ones to provide a more realistic and comprehensive assessment.

%One of the notable strengths of our approach lies in its applicability to any type of digital design, independent of its functionality. Moreover, our method operates solely at the layout level, eliminating the need for prior knowledge about the specific design. This aspect streamlines the security enhancement process, making it more versatile and accessible to a broader range of designs and designers.

%By effectively addressing security concerns during the layout synthesis phase, our approach contributes to a higher level of chip security, safeguarding against potential threats introduced during fabrication or post-fabrication. This has implications for a wide range of applications, particularly in domains where security is paramount, such as financial systems, critical infrastructure, and national security.

%In conclusion, our SALSy approach not only presents a practical and effective solution for enhancing IC security but also provides a versatile and easily implementable method that can significantly benefit various industries and applications.

\bibliographystyle{IEEEtran}
\bibliography{ref}

\begin{thebibliography}{10}
\providecommand{\url}[1]{#1}
\csname url@samestyle\endcsname
\providecommand{\newblock}{\relax}
\providecommand{\bibinfo}[2]{#2}
\providecommand{\BIBentrySTDinterwordspacing}{\spaceskip=0pt\relax}
\providecommand{\BIBentryALTinterwordstretchfactor}{4}
\providecommand{\BIBentryALTinterwordspacing}{\spaceskip=\fontdimen2\font plus
\BIBentryALTinterwordstretchfactor\fontdimen3\font minus
  \fontdimen4\font\relax}
\providecommand{\BIBforeignlanguage}[2]{{%
\expandafter\ifx\csname l@#1\endcsname\relax
\typeout{** WARNING: IEEEtran.bst: No hyphenation pattern has been}%
\typeout{** loaded for the language `#1'. Using the pattern for}%
\typeout{** the default language instead.}%
\else
\language=\csname l@#1\endcsname
\fi
#2}}
\providecommand{\BIBdecl}{\relax}
\BIBdecl

\bibitem{tsmc}
\BIBentryALTinterwordspacing
C.~Ting-Fang, ``Tsmc to triple u.s. chip investment to \$40bn to serve apple,
  others,'' 2022. [Online]. Available:
  \url{https://asia.nikkei.com/Business/Tech/Semiconductors/TSMC-to-triple-U.S.-chip-investment-to-40bn-to-serve-Apple-others}
\BIBentrySTDinterwordspacing

\bibitem{HwSOverview}
W.~Hu, C.-H. Chang, A.~Sengupta, S.~Bhunia, R.~Kastner, and H.~Li, ``An
  overview of hardware security and trust: Threats, countermeasures, and design
  tools,'' \emph{Trans. Comp.-Aided Des. Integ. Circ. Sys.}, vol.~40, no.~6,
  pp. 1010--1038, 2021.

\bibitem{PrimeronHS}
M.~Rostami, F.~Koushanfar, and R.~Karri, ``A primer on hardware security:
  Models, methods, and metrics,'' \emph{Proc. IEEE}, vol. 102, no.~8, pp.
  1283--1295, 2014.

\bibitem{ProtectTrj}
S.~Bhunia, M.~Abramovici, D.~Agrawal, P.~Bradley, M.~S. Hsiao, J.~Plusquellic,
  and M.~Tehranipoor, ``Protection against hardware trojan attacks: Towards a
  comprehensive solution,'' \emph{Des. Test}, pp. 6--17, 2013.

\bibitem{DETERRENT}
V.~Gohil, S.~Patnaik, H.~Guo, D.~Kalathil, and J.~J. Rajendran, ``Deterrent:
  Detecting trojans using reinforcement learning,'' in \emph{Proc. Des. Autom.
  Conf.}, 2022, pp. 697--702.

\bibitem{breakingSilic}
C.~Helfmeier, D.~Nedospasov, C.~Tarnovsky, J.~S. Krissler, C.~Boit, and J.-P.
  Seifert, ``Breaking and entering through the silicon,'' in \emph{Proc. Comp.
  Comm. Sec.}, 2013, p. 733–744.

\bibitem{chipREsurvey}
S.~E. Quadir, J.~Chen, D.~Forte, N.~Asadizanjani, S.~Shahbazmohamadi, L.~Wang,
  J.~Chandy, and M.~Tehranipoor, ``A survey on chip to system reverse
  engineering,'' \emph{J. Emerg. Technol. Comput. Syst.}, 2016.

\bibitem{htsurv}
S.~Bhunia, M.~S. Hsiao, M.~Banga, and S.~Narasimhan, ``Hardware trojan attacks:
  Threat analysis and countermeasures,'' \emph{Proc. IEEE}, pp. 1229--1247,
  2014.

\bibitem{tehrani10}
M.~Tehranipoor and F.~Koushanfar, ``A survey of hardware trojan taxonomy and
  detection,'' \emph{Des. Test}, p. 10–25, 2010.

\bibitem{backdoor}
N.~G. Tsoutsos, C.~Konstantinou, and M.~Maniatakos, ``Advanced techniques for
  designing stealthy hardware trojans,'' in \emph{Proc. Des. Autom. Conf.},
  2014, pp. 1--4.

\bibitem{chipFI}
A.~Barenghi, L.~Breveglieri, I.~Koren, and D.~Naccache, ``Fault injection
  attacks on cryptographic devices: Theory, practice, and countermeasures,''
  \emph{Proc. IEEE}, pp. 3056--3076, 2012.

\bibitem{ChipFI2}
M.~Nagata, ``Exploring fault injection attack resilience of secure ic chips :
  Invited paper,'' in \emph{Proc. Int. Rel. Physics Sym.}, 2022, pp. 1--6.

\bibitem{probing}
H.~Wang, D.~Forte, M.~M. Tehranipoor, and Q.~Shi, ``Probing attacks on
  integrated circuits: Challenges and research opportunities,'' \emph{Des.
  Test}, pp. 63--71, 2017.

\bibitem{tehraniFSP}
H.~Wang, Q.~Shi, A.~Nahiyan, D.~Forte, and M.~M. Tehranipoor, ``A physical
  design flow against front-side probing attacks by internal shielding,''
  \emph{Trans. Comp.-Aided Des. Integ. Circ. Sys.}, pp. 2152--2165, 2020.

\bibitem{eslami23}
M.~Eslami, J.~Knechtel, O.~Sinanoglu, R.~Karri, and S.~Pagliarini,
  ``Benchmarking advanced security closure of physical layouts: Ispd 2023
  contest,'' in \emph{Proc. Int. Symp. Phys. Des.}, 2023, p. 256–264.

\bibitem{TrjClass}
R.~Karri, J.~Rajendran, K.~Rosenfeld, and M.~Tehranipoor, ``Trustworthy
  hardware: Identifying and classifying hardware trojans,'' \emph{Computer},
  pp. 39--46, 2010.

\bibitem{trojanLessons}
K.~Xiao, D.~Forte, Y.~Jin, R.~Karri, S.~Bhunia, and M.~Tehranipoor, ``Hardware
  trojans: Lessons learned after one decade of research,'' \emph{Trans. Des.
  Autom. Elec. Sys.}, 2016.

\bibitem{Alex}
A.~Hepp, T.~Perez, S.~Pagliarini, and G.~Sigl, ``A pragmatic methodology for
  blind hardware trojan insertion in finalized layouts,'' in \emph{Proc. Int.
  Conf. Comp.-Aided Des.}, 2022.

\bibitem{tiago_tcad}
T.~D. Perez and S.~Pagliarini, ``Hardware trojan insertion in finalized
  layouts: From methodology to a silicon demonstration,'' \emph{Trans.
  Comp.-Aided Des. Integ. Circ. Sys.}, vol.~42, no.~7, pp. 2094--2107, 2023.

\bibitem{laserFI}
B.~Selmke, J.~Heyszl, and G.~Sigl, ``Attack on a {DFA} protected {AES} by
  simultaneous laser fault injections,'' in \emph{Proc. Worksh. Fault Diag.
  Tol. Cryptogr.}, 2016, pp. 36--46.

\bibitem{laserFI2}
R.~A.~C. Viera, P.~Maurine, J.-M. Dutertre, and R.~Possamai~Bastos,
  ``Simulation and experimental demonstration of the importance of ir-drops
  during laser fault injection,'' \emph{Trans. Comp.-Aided Des. Integ. Circ.
  Sys.}, pp. 1231--1244, 2020.

\bibitem{backside}
C.~Boit, S.~Tajik, P.~Scholz, E.~Amini, A.~Beyreuther, H.~Lohrke, and J.~P.
  Seifert, ``From {IC} debug to hardware security risk: The power of backside
  access and optical interaction,'' in \emph{Proc. Int. Symp. Physical Failure
  Analys. IC}, 2016, pp. 365--369.

\bibitem{second}
F.~Wang, Q.~Wang, B.~Fu, S.~Jiang, X.~Zhang, L.~Alrahis, O.~Sinanoglu,
  J.~Knechtel, T.-Y. Ho, and E.~F. Young, ``Security closure of ic layouts
  against hardware trojans,'' in \emph{Proc. Int. Symp. Phys. Des.}, 2023, p.
  229–237.

\bibitem{bisa}
K.~Xiao and M.~Tehranipoor, ``Bisa: Built-in self-authentication for preventing
  hardware trojan insertion,'' in \emph{Proc. Int. Symp. Hardw.-Orient. Sec.
  Trust}, 2013, pp. 45--50.

\bibitem{papa}
P.-S. Ba, S.~Dupuis, M.~Palanichamy, M.-L. Flottes, G.~Di~Natale, and
  B.~Rouzeyre, ``Hardware trust through layout filling: A hardware trojan
  prevention technique,'' in \emph{Proc. Comp. Soc. Symp. VLSI}, 2016, pp.
  254--259.

\bibitem{JohannICCAD}
J.~Knechtel, J.~Gopinath, J.~Bhandari, M.~Ashraf, H.~Amrouch, S.~Borkar, S.-K.
  Lim, O.~Sinanoglu, and R.~Karri, ``Security closure of physical layouts iccad
  special session paper,'' in \emph{Proc. Int. Conf. Comp.-Aided Des.}, 2021,
  pp. 1--9.

\bibitem{ASSURER}
G.~Guo, H.~You, Z.~Tang, B.~Li, C.~Li, and X.~Zhang, ``Assurer: A ppa-friendly
  security closure framework for physical design,'' in \emph{Proc. Asia South
  Pac. Des. Autom. Conf.}, 2023, p. 504–509.

\bibitem{ter}
T.~Trippel, K.~G. Shin, K.~B. Bush, and M.~Hicks, ``T-ter: Defeating a2 trojans
  with targeted tamper-evident routing,'' in \emph{Proc. of the 2023 ACM Asia
  Conf. on Comp. and Comm. Sec.}, 2023, p. 746–759.

\bibitem{contest}
J.~Knechtel, J.~Gopinath, M.~Ashraf, J.~Bhandari, O.~Sinanoglu, and R.~Karri,
  ``Benchmarking security closure of physical layouts: {ISPD} 2022 contest,''
  in \emph{Proc. Int. Symp. Phys. Des.}, 2022, p. 221–228.

\bibitem{benchmarks}
\BIBentryALTinterwordspacing
T.~Sugawara, N.~Homma, T.~Aoki, and A.~Satoh, ``Asic performance comparison for
  the iso standard block ciphers,'' in \emph{JWIS}, 2007. [Online]. Available:
  \url{http://www.aoki.ecei.tohoku.ac.jp/crypto/web/cores.html}
\BIBentrySTDinterwordspacing

\bibitem{benchmarks1}
\BIBentryALTinterwordspacing
O.~Girard~et al., ``openmsp430 at opencores.org,'' 2021. [Online]. Available:
  \url{https://opencores.org/projects/openmsp430}
\BIBentrySTDinterwordspacing

\bibitem{benchmarks2}
\BIBentryALTinterwordspacing
M.~Hicks~et al., ``Mit-ll common evaluation platform (cep) at github.com,''
  2021. [Online]. Available: \url{https://github.com/mit-ll/CEP}
\BIBentrySTDinterwordspacing

\bibitem{nangate}
\BIBentryALTinterwordspacing
``Nangate freepdk45 open cell library.'' [Online]. Available:
  \url{http://www.nangate.com/?page_id=2325}
\BIBentrySTDinterwordspacing

\bibitem{github}
\BIBentryALTinterwordspacing
``Salsy repository.'' [Online]. Available:
  \url{https://github.com/Centre-for-Hardware-Security/SALSy}
\BIBentrySTDinterwordspacing

\bibitem{A2Trojan}
K.~Yang, M.~Hicks, Q.~Dong, T.~Austin, and D.~Sylvester, ``A2: Analog malicious
  hardware,'' in \emph{Proc. Symp. Sec. Priv.}, 2016, pp. 18--37.

\end{thebibliography}

\begin{IEEEbiography}[{\includegraphics[width=1in,height=1.25in,clip,keepaspectratio]{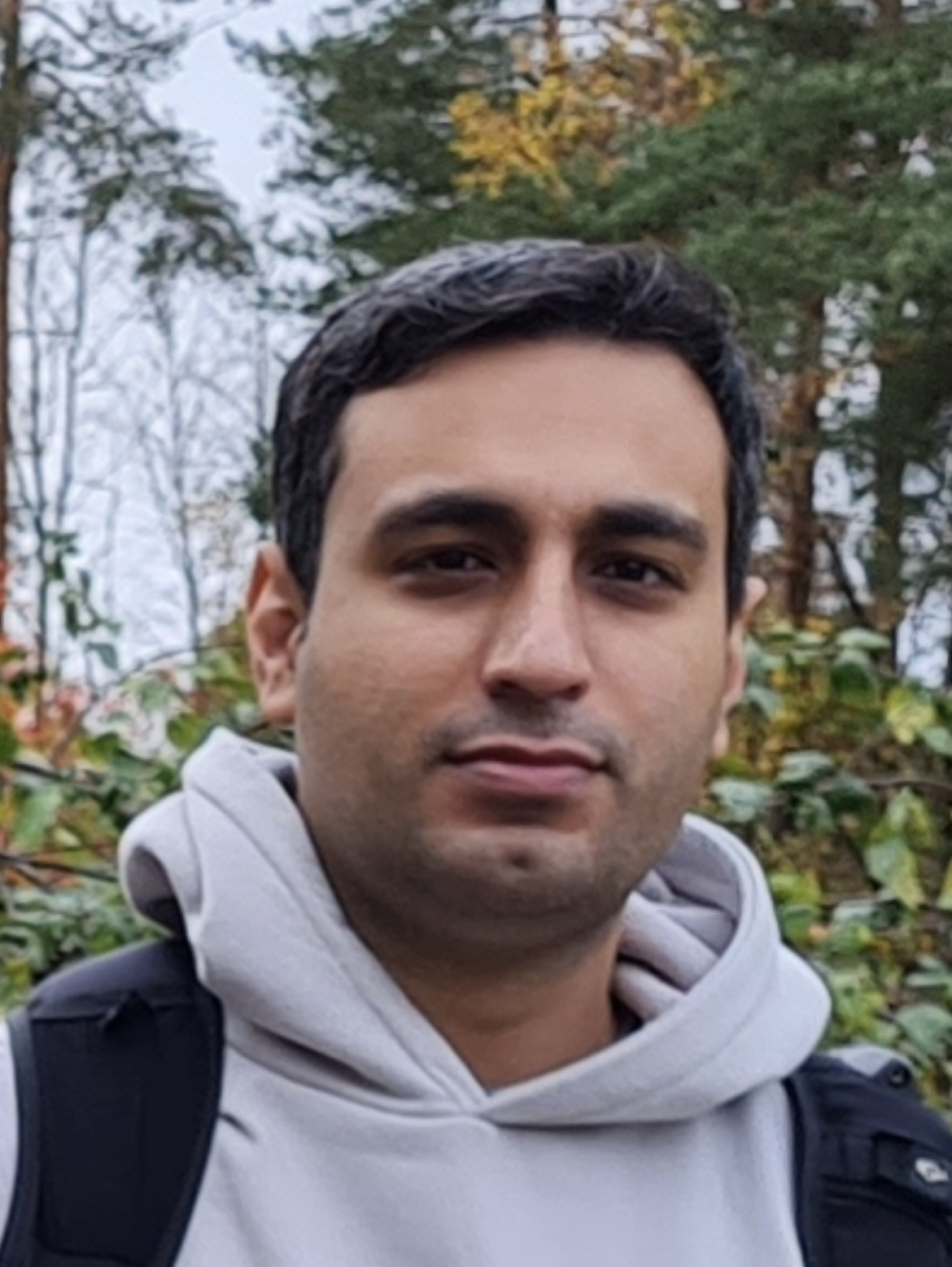}}]{Mohammad Eslami}
received his M.S. degree in computer engineering from the Shahid Bahonar University of Kerman, Kerman, Iran, in 2018. Currently, he is pursuing his doctoral studies at the Centre for Hardware Security, Tallinn University of Technology (TalTech), Tallinn, Estonia. 

His research interests primarily revolve around hardware security, with a particular focus on physical design automation and secure ASIC design.
\end{IEEEbiography}

\vskip -2\baselineskip plus -1fil

\begin{IEEEbiography}[{\includegraphics[width=1in,height=1.25in,clip,keepaspectratio]{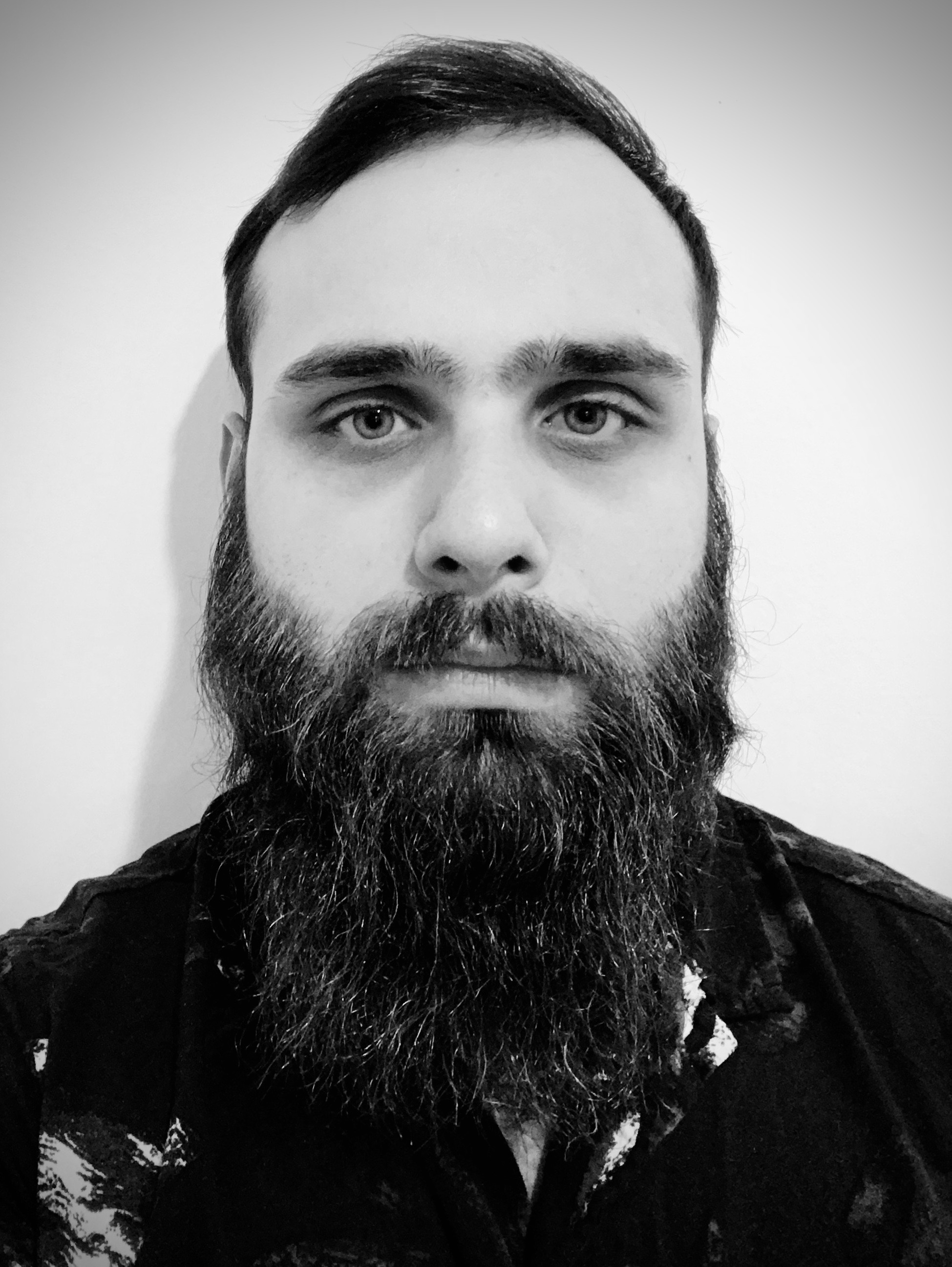}}]{Tiago D. Perez} received the M.S. degree in electric engineering from the University of Campinas, S\~ao Paulo, Brazil, in 2019. He has also received the Ph.D. degree from Tallinn University of Technology (TalTech), Tallinn, Estonia. 

From 2014 to 2019, he was a Digital Designer Engineer with Eldorado Research Institute, S\~ao Paulo, Brazil. His research interests include digital signal processing, telecommunication systems, IC design, and hardware security.
\end{IEEEbiography}

\vskip -2\baselineskip plus -1fil

\begin{IEEEbiography}[{\includegraphics[width=1in,height=1.25in,clip,keepaspectratio]{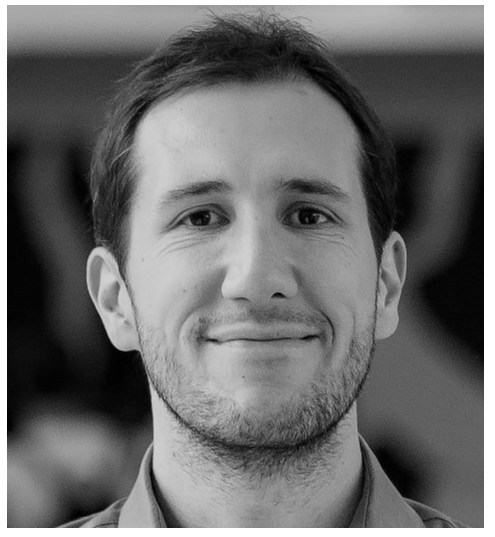}}]{Samuel Pagliarini}
(M'14) received the PhD degree from Telecom ParisTech, Paris, France, in 2013. 

He has held research positions with the University of Bristol, Bristol, UK, and with Carnegie Mellon University, Pittsburgh, PA, USA. He is currently a Professor of Hardware Security with Tallinn University of Technology (TalTech) in Tallinn, Estonia where he leads the Centre for Hardware Security. His current research interests include many facets of digital circuit design, with a focus on circuit reliability, dependability, and hardware trustworthiness.

\end{IEEEbiography}

\vfill

\end{document}